\documentclass{aa}

\usepackage{graphicx}
\usepackage{txfonts}
\usepackage{hyperref}
\usepackage{epstopdf}
\usepackage{xcolor}

\newcommand{\R}{\textsf{R}}

\begin{document} 

   \title{Unsupervised classification of SDSS galaxy spectra}


   \author{D. Fraix-Burnet
          \inst{1}
          \and
          C. Bouveyron\inst{2}
          \and 
          J. Moultaka\inst{3}
          }

   \institute{Univ. Grenoble Alpes, CNRS, IPAG, Grenoble, France \\ \email{didier.fraix-burnet@univ-grenoble-alpes.fr}
         \and Université Côte d'Azur, Inria, CNRS, LJAD, Maasai, Nice, France\\
             \email{charles.bouveyron@univ-cotedazur.fr}
           \and
           IRAP, Université de Toulouse, CNRS, CNES, UPS, 14, avenue Edouard Belin, F-31400 Toulouse, France\\
           \email{jihane.moultaka@irap.omp.eu}
             }

   \date{Received July 31, 2019; accepted September 15, 2019}

 \abstract
 {Defining templates of galaxy spectra is useful to quickly characterise new observations and organise databases from surveys. These templates are usually built from a pre-defined classification based on other criteria.}
 {We present an unsupervised classification of 702248 spectra of galaxies and quasars with redshifts smaller than 0.25 that were retrieved from the Sloan Digital Sky Survey (SDSS) database, release 7. }
 {The spectra were first corrected for redshift, then wavelet-filtered to reduce the noise, and finally binned to obtain about 1437 wavelengths per spectrum. The unsupervised clustering  algorithm Fisher-EM, relying on a discriminative latent mixture model, was applied on these corrected spectra. The full set and several subsets of 100000 and 300000 spectra were analysed.}
 {The optimum number of classes given by a penalised likelihood criterion is 86 classes, of which the 37 most populated gather 99\% of the sample. These classes are established from a subset of 302214 spectra. Using several cross-validation techniques we find that this classification agrees with the results obtained on the other subsets with an average misclassification error of about 15\%. The large number of very small classes tends to increase this error rate. In this paper, we do an initial quick comparison of our classes with literature templates. 
 }
 {This is the first time that an automatic, objective and robust unsupervised classification is established on such a large number of galaxy spectra. The mean spectra of the classes can be used as templates for a large majority of galaxies in our Universe.}

   \keywords{Methods: data analysis --
                Methods: statistical --
                Galaxies: statistics --
                Galaxies: general --
                Techniques: spectroscopic
               }

   \maketitle
%

\section{Introduction}
\label{introduction}

There is a long recognised need for templates of typical galaxy spectra to quickly characterise new objects. For instance, such templates are used in automatic classification in databases such as the {Sloan Digital Sky Survey (SDSS)}, in which a $\chi^2$ minimisation is used to separate galaxies from {quasars (QSOs)} and stars, followed by a categorisation into starbursts, star-forming galaxies and {active galactic nuclei (AGN)} using line ratios\footnote{\url{https://www.sdss.org/dr12/spectro/catalogs/\#Objectinformation}}.
		
Templates can be built by stacking spectra to reduce the noise that can make some features disappear \citep{Dobos2012}. This strategy obviously requires perfectly identical spectra that can be obtained by visual inspection, by a classification made from some general or specific properties of the galaxies (e.g. line ratios, continuum, morphology, and model fitting of stellar population), or by machine learning tools.

\citet{Kennicutt1992} proposed an atlas of spectra classified according to the morphology of the galaxies. This kind of approach makes the assumption of a perfect correspondence between the two pieces of information and is insufficient to characterise the informative wealth of the spectra.

More in-depth classifications of spectra are driven by several spectral features such as emission lines or their combinations with colours, nuclear, and star formation activity and morphology. This results in a somewhat complicated scheme with 38 classes \citep{Dobos2012} or in a rather general classification with only 6 classes  \citep{Wang2018}.

In all cases above, the classification is pre-defined by physics. In \citet{Dobos2012}, the template family is not a classification since a given spectrum can belong to several categories, for instance one for colour and one for nuclear activity or star formation. Using these templates as a library to generate mock catalogues through interpolation, or to be fitted by models of galaxy composition and evolution may be fine, but characterising a new spectrum might become problematic. Another difficulty in this classification comes from the sharp arbitrary cuts necessary to distinguish two spectra. For instance, {\citet{Dobos2012} classified only those with $H\alpha$ measured at the signal-to-noise ratio $S/N\ge 3$}  as star-forming galaxies.

\citet{Wang2018} performed a classification of {Large Sky Area Multi-Object Fibre Spectroscopic Telescope (LAMOST)} spectra not observed in SDSS as used in \citet{Dobos2012} or our present paper. Their classification is based only on spectral features measured beforehand and using classical diagnostic {Baldwin-Phillips-Terlevich (BPT)} diagrams \citep{Baldwin1981}, thereby explaining that only six classes can be defined. \citet{Wang2018} first compare the distribution of galaxies in their classes with that from \citet{Dobos2012} and find that the differences are explained by the different cuts used to categorise the classes. \citet{Wang2018} then compare the median composite spectra themselves simply using the a priori correspondence of the classes. They conclude that the continua match well but not the emission lines. 

The main drawback of most of these studies is that they mainly rely on a visual inspection. This limits the sample size used to establish the templates and thus their representativeness of the galaxy diversity. New methods should make use of the information contained in the spectra themselves to become both automatic and objective, and machine learning tools of unsupervised classification (clustering) are well suited for this \citep{Fraix-Burnet2015}.

The goal of unsupervised classification is to detect structures in the data space from which clusters of similar spectra can be identified. The average spectrum of a cluster defines a typical template that is dictated only by the data. The physics comes afterwards, by inverse procedures or model fitting to characterise the underlying stellar population and the ionisation conditions within the corresponding galaxy \citep[e.g.][]{Moultaka2004,Noll2009,Comparat2017}. Then, these clusters can be considered as classes.

Principal component analysis (PCA) defines an orthogonal base of spectra from which templates can be constructed. For instance, \citet{Marchetti2013} find that three {principal components} are sufficient to describe the variance of their sample, and the coefficients (loadings) of each spectra are used to perform an unsupervised k-means classification. However, the variance axes are not necessarily the discriminant axes \citep{Chang1983}.

Other ways of classifying spectra {are} to use the embedded information, such as the independent component analysis \citep[ICA, ][]{Lu2006} or t-distributed stochastic neighbour embedding (t-SNE) reduction of spectral information which, to our knowledge, has only been applied on stellar spectra \citep{Traven2016}. Some efficient tools can be designed for specific purposes, such as the detection of outliers \citep{Baron2017}.

So far, no robust unsupervised classification of galaxy spectra has been proposed. An attempt with raw spectra and the k-means approach was performed by \citet{SanchezAlmeida2010} but this result is disputed by \citep{De2016}. Indeed, this technique does not perform well in high dimension, when the classes are not balanced and well separated in parameter space.

A more powerful tool is the Gaussian mixture model \citep[GMM, ][]{Bouveyron2019} approach which fits a multivariate Gaussian to each cluster \cite[e.g.][]{Souza2017}. Mixture models are particularly popular in the context of clustering since they facilitate fitting to various situations and to automatically find the appropriate number of clusters in the data at hand. In addition, in the case of a high-dimensional space, subspace clustering usually yields more robust results that can be both justified from the statistical point of view (curse of dimensionality) and the physical argument that not all parameters are discriminant. This is especially true for spectra because of the noise and many uninformative wavelengths.

In this paper, we use a GMM approach implemented in the Fisher-EM algorithm \citep{Bouveyron2012a}. It is a subspace clustering algorithm, based on a latent mixture model that fits the data into a low-dimensional discriminative subspace. Fisher-EM uses a modified version of the expectation-maximisation (EM) algorithm for model inference and introduces a Fisher criterion to optimise the quality of the clustering. It has been used only once in astrophysics up to now, on a data set of 100000 galaxies described by 12 parameters \citep{Siudek2018a}. In the present paper, we apply Fisher-EM on a larger sample of 702248 spectra (with 1437 wavelengths) from the SDSS.

The paper is organised as follows. The data are described in Sect.~\ref{data}. The Fisher-EM algorithm and its application on our large data set are detailed in Sect.~\ref{method}. The results are given in Sect.~\ref{results} and an estimation of the statistical robustness of our classification is presented in Sect.~\ref{robustness}. To get a first idea of the physical meaning of our classes, we present a comparison with the existing atlases of \citet{Kennicutt1992, Dobos2012, Wang2018}
{ in Sect.~\ref{discussion}.}
A discussion and a conclusion are given in Sects.~\ref{discussion} and \ref{conclusions} respectively.

\section{The data}
\label{data}

In this work, we use the same data set as studied by \citet{SanchezAlmeida2010} and \citet{De2016}. Thus, we retrieved the spectra of 702248 galaxies and QSOs (with redshift smaller than 0.25) from the SDSS database, release 7\footnote{\url{http://www.sdss.org/dr7/}}. {For consistency with the two previous studies, we chose not to correct the spectra for galactic extinction. Apart from a philosophical justification explained in \citet{SanchezAlmeida2010}, the correction would be small (median extinction \textit{g-r} of 0.03 for our sample) and unavoidably somewhat uncertain, bringing an additional variance (see a further comment in Sect.~\ref{description}).} The raw spectra have 3850 points in the wavelength range imposed by the instrument, $\lambda = 3800$ to $9250$~\AA. The spacing is uniform in resolution ($\delta\lambda/\lambda = 1/4342$). Taking the redshift into account, the range common to all the spectra goes  from 3806 to 7371~\AA\ with 2874 wavelengths.
We then corrected the spectra for the redshift, as in \citet{De2016}, using the Shannon criterion to preserve the shape of the spectral lines. For this purpose, we doubled the sampling of the spectra beforehand, generating 5748 points for each spectrum. The resulting spectra were then normalised by dividing each spectrum by its average value between 4300 and 5000~\AA. 

Contrary to \citet{SanchezAlmeida2010} and \citet{De2016}, we did not select a priori bands supposedly containing relevant physical information to reduce the size of data array.  Instead, we first applied a noise reduction through wavelet filtering followed by a binning by a factor of 4 to obtain spectra with 1437 wavelengths, which is a number comparable with that used in the two previous works. In this process, we keep most of the information with the shapes of the lines being preserved (see Appendix~\ref{noise}) as well as full objectivity. These computations were done using algorithms under the \R\ environment\footnote{\url{https://www.r-project.org/}, algorithms \textit{wd, threshold} and \textit{wr} from the package \textit{wavethresh}, and \textit{binning} from the package \textit{prospectr}.}.

\section{The Fisher-EM algorithm}
\label{method}

\subsection{Mathematical overview of the Fisher-EM algorithm}
\label{fem}

The Fisher-EM algorithm relies on a statistical model, named the discriminative latent mixture (DLM) model, and uses a modified version of the EM algorithm for model inference. We provide hereafter a short mathematical overview of both the model and the inference algorithm.

The DLM model assumes that $\{y_{1},\dots,y_{n}\}\in\mathbb{R}^{p}$ denotes the data set of
$n$ observations that we want to cluster into $K$ homogeneous
groups, that is adjoin to each observation $y_{i}$ a value $z_{i}\in\{1,\dots,K\}$
where $z_{i}=k$ indicates that the observation $y_{i}$ belongs to
the $k$th group. On the one hand, let us assume that $\{y_{1},\dots,y_{n}\}$
are independent observed realisations of a random vector $Y\in\mathbb{R}^{p}$
and that $\{z_{1},\dots,z_{n}\}$ are also independent realisations
of a random variable $Z\in\{1,\dots,K\}$. On the other hand, let
$\mathbb{E}\subset\mathbb{R}^{p}$ denote a latent space assumed to
be the most discriminative subspace of dimension $d\leq K-1$ such
that $\mathbf{0}\in\mathbb{E}$. 
Moreover, let $\{x_{1},\dots,x_{n}\}\in\mathbb{E}$
denote the actual data, described in the latent space~$\mathbb{E}$
of dimension $d$, which are in addition presumed to be independent
realisations of an unobserved random vector $X\in\mathbb{E}$. Finally,
the observed variable $Y\in\mathbb{R}^{p}$ and the latent variable
$X\in\mathbb{E}$ are assumed to be linked through a linear transformation as follows:
\begin{equation}
Y=UX+\varepsilon,\label{eq:linear_relationship}
\end{equation}
where $U$ is a $p\times d$ orthonormal matrix common to the $K$
groups and satisfying $U^{t}U=\mathbf{I}_{d}$. The $p$-dimensional
random vector $\varepsilon$ stands for the noise term that models
the non-discriminative information and is assumed to be distributed
according to a centred Gaussian density function with a covariance
matrix $\Psi$ ($\varepsilon_{k}\sim\mathcal{N}(0,\Psi_{k})$). Besides, within
the latent space, $X$ is assumed, conditionally to $Z=k$, to be
Gaussian, where
\begin{equation}
X_{|Z=k}\sim\mathcal{N}(\mu_{k},\Sigma_{k})
\end{equation}
and $\mu_{k}\in\mathbb{R}^{d}$ and $\mbox{\ensuremath{\Sigma}}_{k}\in\mathbb{R}^{d\times d}$
are the mean vector and the covariance matrix of the
$k$th group, respectively. Given these distribution assumptions and according to
equation~(\ref{eq:linear_relationship}), 
\begin{equation}
Y_{|X,Z=k}\sim\mathcal{N}(UX,\Psi_{k}),\label{eq:Y}
\end{equation}
and its marginal distribution is therefore a mixture of Gaussians as follows:
\begin{equation}
f(y)=\sum_{k=1}^{K}\pi_{k}\phi(y;m_{k},S_{k}),\label{eq:GMM}
\end{equation}
where $\pi_{k}$ is the mixing proportion of the $k$th group and
$\phi(.;m_{k},S_{k})$ denotes the multivariate Gaussian density function
parametrised by the mean vector $m_{k}=U\mu_{k}$ and the covariance
matrix $S_{k}=U\Sigma_{k}U^{t}+\Psi_{k}$ of the $k$th group. Furthermore,
we define the $p\times p$ matrix $W=[U,V]$ such that $W^{t}W=WW^{t}=\mathbf{I}_{p}$,
where the $(p-d)\times p$ matrix $V$ is an orthogonal complement
of $U$. Finally, the noise covariance matrix $\Psi_{k}$ is assumed to
satisfy the conditions $V^{t}\Psi_{k} V=\beta_{k}\mathbf{I}_{p-d}$ and $U^{t}\Psi_{k} U=\mathbf{0}_{d}$,
such that $\Delta_{k}=W^{t}S_{k}W$ has the following form: $$\Delta_k=\left(  \begin{array}{c@{}c} \begin{array}{|ccc|}\hline ~~ & ~~ & ~~ \\  & \Sigma_k &  \\  & & \\ \hline \end{array} & \mathbf{0}\\ \mathbf{0} &  \begin{array}{|cccc|}\hline \beta_{k} & & & 0\\ & \ddots & &\\  & & \ddots &\\ 0 & & & \beta_{k}\\ \hline \end{array} \end{array}\right)  \begin{array}{cc} \left.\begin{array}{c} \\\\\\\end{array} \right\}  & d \leq K-1\vspace{1.5ex}\\ \left.\begin{array}{c} \\\\\\\\\end{array}\right\}  & (p-d)\end{array}$$ These
last conditions imply that the discriminative and the non-discriminative
subspaces are orthogonal, which suggests in practice that all the
relevant clustering information remains in the latent subspace. From a practical point of view, $\beta_{k}$ models the variance of the non-discriminative noise of the data. The whole
model is referred to by DLM$_{[\Sigma_{k}\beta_{k}]}$ in \cite{Bouveyron2012a}.

Several other models can be obtained from the DLM$_{[\Sigma_{k}\beta_{k}]}$
model by relaxing or adding constraints on model parameters. For instance, the covariance matrices $\Sigma_{1}...\Sigma_{K}$ and the noise parameters $\beta_{1} ... \beta_{K}$ can be assumed to be common to all classes (hence independent on $k$). This is the statistical model DLM$_{[\Sigma\beta]}$ (referred to as DB in the \R\ package \textit{FisherEM}). A list
of the $12$ different DLM models is given in Table~1 in \cite{Bouveyron2012a}.
Such a family simultaneously yields very parsimonious models and allows us to carry out a fit in various situations.

An estimation procedure, called the Fisher-EM algorithm, is also proposed
in \cite{Bouveyron2012a} to estimate the discriminative
space and parameters of the mixture model. This algorithm is based
on the EM algorithm to which an additional step is introduced between
the E and M-step. This additional step, named F-step, aims to
compute the projection matrix $U$ whose columns span the discriminative
latent space. The Fisher-EM algorithm has therefore the following
form, at iteration~$q$:
\begin{itemize}
\item E-step: This step computes the posterior probabilities $t_{ik}^{(q)}$ that
the observations belong to the $K$ groups using the following update
formula:
\begin{equation}
t_{ik}^{(q)}=\hat{\pi}_{k}^{(q-1)}\phi(y_{i},\hat{\theta}_{k}^{(q-1)})/\sum_{\ell=1}^{K}\hat{\pi}_{\ell}^{(q-1)}\phi(y_{i},\hat{\theta}_{\ell}^{(q-1)}),\label{eq:t_ik}
\end{equation}
where $\hat{\theta}_{k}=\{\hat{\mu}_{k},\hat{\Sigma}_{k},\hat{\beta}_{k},\hat{U}\}$. 
\item F-step: This step estimates, conditionally to the posterior probabilities,
the orientation matrix $U^{(q)}$ of the discriminative latent space
by maximising Fisher's criterion \citep[defined by the ratio of the between-class variance to the within-class variance,][]{Fisher1936,Fukunaga2013}
under the following orthonormality constraints:
\begin{eqnarray}
\hat{U}^{(q)} & = & \max_{U}\quad\mathrm{trace}\left((U^{t}SU)^{-1}U^{t}S_{B}^{(q)}U\right),\nonumber \\
 &  & \text{w.r.t.}\quad U^{t}U=\mathbf{I}_{d},\label{eq:Opti_Fisher}
\end{eqnarray}
where $S$ stands for the covariance matrix of the whole data set and
$S_{B}^{(q)}$, defined as follows: 
\begin{equation}
S_{B}^{(q)}=\frac{1}{n}\sum_{k=1}^{K}n_{k}^{(q)}(m_{k}^{(q)}-\bar{y})(m_{k}^{(q)}-\bar{y})^{t},\label{eq:Sb_soft}
\end{equation}
denotes the soft between-covariance matrix with $n_{k}^{(q)}=\sum_{i=1}^{n}t_{ik}^{(q)}$,
$m_{k}^{(q)}=\sum_{i=1}^{n}t_{ik}^{(q)}y_{i}/n_{k}^{(q)}$ and $\bar{y}=\sum_{i=1}^{n}y_{i}/n$.
\item M-step: This third step estimates the parameters of the mixture model in the
latent subspace by maximising the conditional expectation of the complete
log-likelihood.
\end{itemize}

The Fisher-EM procedure iteratively updates the posterior probabilities
and the parameters until the Aitken criterion is satisfied (see paragraph~4.5
of \cite{Bouveyron2012a}). The convergence properties of the Fisher-EM
algorithm were studied in \cite{Bouveyron2012b}. Finally, since the latent subspace has a low dimension and is common
to all groups, the clustered data can be easily visualised by projecting
them into the estimated latent subspace.

The choice of the best statistical DLM model and the optimum number of clusters depends on the data and are estimated with three likelihood criteria: the classical Akaike information criterion (AIC), the popular Bayesian information criterion (BIC) and the integrated completed likelihood (ICL). These criteria all penalise the likelihood by the number of parameters of the statistical model, but the BIC also includes the number of observations while the ICL in addition favours well separated statistical models \citep{Biernacki2000,Girard2016}. Hence, the maximum value of ICL determines the best statistical model and the optimum number $K$ of classes in the analyses performed in the present study.

The Fisher-EM algorithm is implemented in the eponym package for \R. For the present study, we used the version 1.5.1 of this package.

\subsection{Differences between k-means and Fisher-EM}
\label{KmeansvsFEM}

The clustering algorithm k-means is relatively standard and easy to use. However, this algorithm has severe limitations that triggered the development of generalisations, Fisher-EM being one of the most recent and complete. The statistical differences can be summarised as follows:

	1) K-means assumes that all classes can be fit with the same pattern, that is a spherical Gaussian. The Fisher-EM algorithm belongs to the class of GMMs that releases this strong assumption.
	
	2)  K-means is a hard clustering algorithm: each observation belongs to one class and only one. Fisher-EM, which is based on the EM algorithm, is soft in the sense that each observation is associated with a vector of probabilities for all classes. In Fisher-EM, an observation is attributed to a class if the corresponding probability is higher than 0.5.
	
	3) K-means works in the full data space, while Fisher-EM looks for a subspace in which the classes are more conspicuous, and thus makes a selection of the most pertinent dimensions.
	
	4) K-means uses only the Euclidean distance to define classes, while Fisher-EM takes into account the variance and optimises the compactness of classes.
	
	5) {K-means always yields a solution, while Fisher-EM, as all EM algorithms, may not be able to converge to a local maximum \citep{Wu1983,Bouveyron2012b}.}

{All possible classes of galaxy spectra can certainly not be represented by either spherical or identical distributions in the data space, and many wavelengths bring either no information or just noise. These two points favour the use of subspace and GMM-based clustering algorithms for large samples of astrophysical spectra. }

\subsection{Application to the large data set of SDSS spectra}
\label{choice}

The selection of the best statistical model and the optimum number of classes requires many computations. Since our data set is very large, it is inconvenient in practice to use the full sample at once. With a low number of classes, say K=3, the analysis with Fisher-EM takes about 30 hours CPU time for 702248 spectra. But for K=50, on a smaller sample of 300000 spectra, it takes several hundreds of hours. Considering that the computation has to be repeated for each of the 12 statistical models and for many values of K to find the maximum ICL value, splitting the sample into subsets makes the analyses manageable and allows for robustness estimation.

\begin{figure}
	\includegraphics[width=\linewidth]{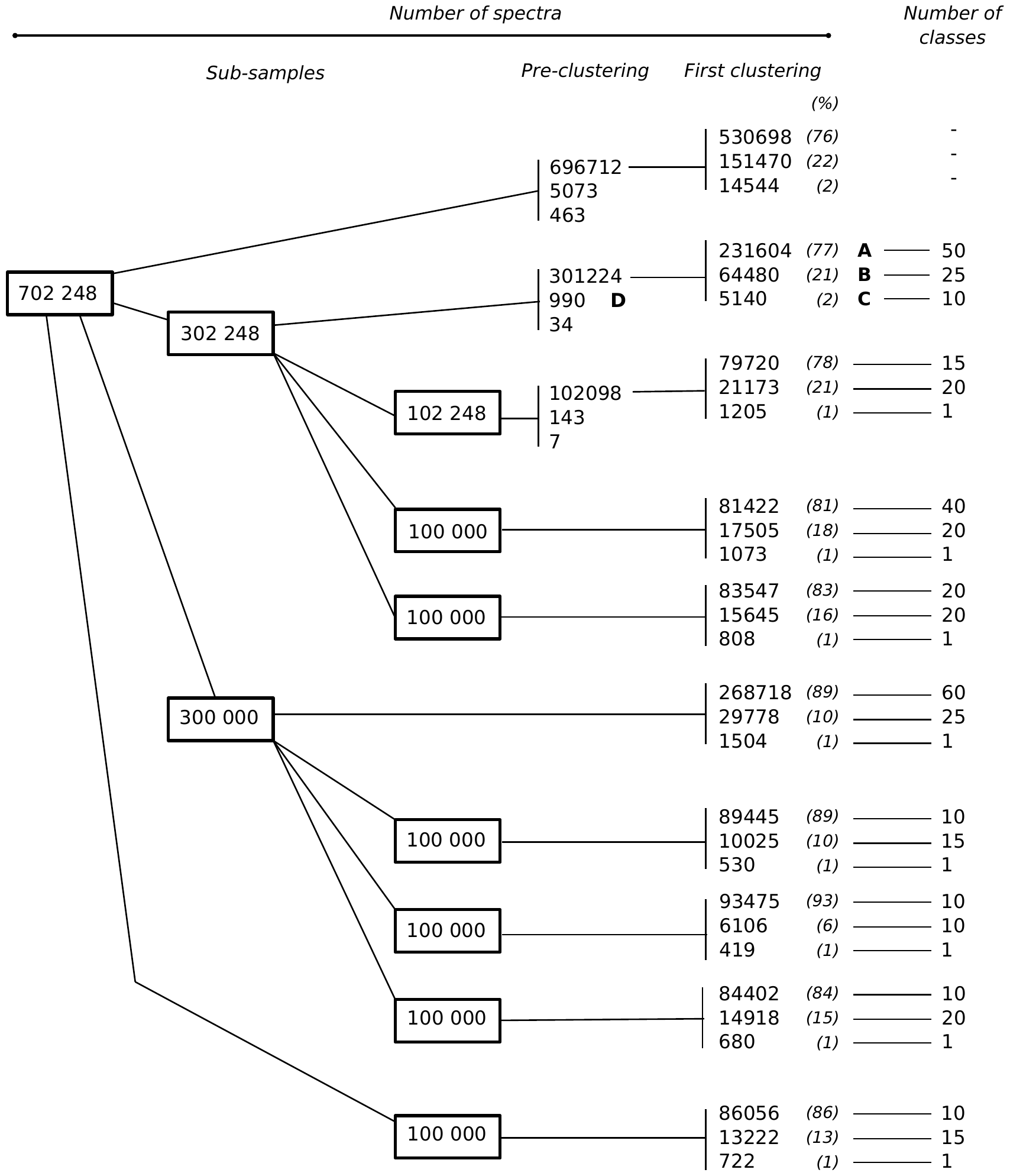}
	\caption{Summary of all the clustering analyses performed on different sub-samples. The size of each sample is given in the boxes, and the sizes of the different classes are given in the first two columns to the right, with the percentages given in parentheses. The optimum number of classes is given in the last column to the right. The three upmost samples required a pre-clustering with K=3 before the first clustering also with K=3 on the largest class gathering more than 99\% of the spectra. The analysis of the full sample (702248 spectra) was not pursued after the second pre-classification for excessive computation time. The four main classes A, B, C and D are indicated in boldface.}
	\label{fig:schemeanalyses}
\end{figure}

We thus divided the full sample into several subsets that have been analysed separately and for different purposes. We are aware that this approach results in a somewhat complicated scheme, but in this exploring stage of the application of Fisher-EM on such a large astronomical sample, we believe it is essential to perform thorough and robust analyses to be confident with the resulting classification and to suggest tractable strategies to tackle even larger samples. These subsets were obtained simply by arbitrarily splitting the full sample assuming that the order within the SDSS database would provide enough randomisation of the galaxy diversity. We summarise our approach in Fig.~\ref{fig:schemeanalyses} as follows:

		i) The full sample with 702248 spectra was only partially analysed, because the very good agreement between all our analyses did not seem to justify the large computation time that it would have required (Sect.~\ref{robustness}).
		
		ii) Two sub-samples of 300000 and 302248 spectra were fully analysed. The latter yields our final classification (Sect.~\ref{results}), while the former was used for comparison  (Sect.~\ref{robustness}).
		
		iii) Seven sub-samples of $\simeq 100000$ spectra were used for a first exploration of the data set and a search for the best statistical model (see below). They were also used as a robustness assessment (Sect.~\ref{robustness}).

With the 100000 subsets, the best statistical model is found to be DLM$_{[\Sigma\beta]}$ (see Sect.~\ref{fem}). Since almost all the other models did not even converge towards any classification, we adopted this statistical model in the all the analyses performed for this paper.

We also found that when increasing the number of classes $K$, the number of very small classes increases while the two to three larger classes remain nearly untouched. As an example, for a sample of 100000 spectra and K=2, the distribution of the number of spectra in the two classes is 94654, 5346, for K=3 we get 86056, 13222, 722, and for K=50 we get 74007, 22129, 1324, 757, 669, 204... with ten classes that only have one spectrum. {This behaviour was also found on the larger subsets.}

We thus decided to first cluster each subset into three classes, and then apply the Fisher-EM algorithm separately on the two largest classes {to obtain a more refined classification}.  The sub-clustering process can be pursued as much as required by the purpose of the classification and its desired granularity. 
This {procedure is similar to} the "canopy" approach proposed in \citet{De2016} in which a first rough clustering is performed before a more refined analysis. Because the Fisher-EM algorithm is much more precise and discriminant than the k-means tool used in \citet{De2016}, we often end up with a few classes that have only a very small number of spectra even down to one single spectrum.

{This procedure is depicted in Fig.~\ref{fig:schemeanalyses} in the column labelled "First clustering" (hence with K=3) while the last column gives the optimum number of clusters obtained by the sub-clustering analyses of the three first clusters. However a supplementary step with K=3, shown by the column labelled "Pre-clustering" in Fig.~\ref{fig:schemeanalyses} was necessary for the three samples with 70248, 302248 and 102248, thus containing the same 102248 spectra. This is explained in Sect.~\ref{othersubsets}.}

One can wonder why we need the sub-clustering of classes and why we do not find initially the total number of classes. This is because the clustering {made by Fisher-EM} is defined in a latent subspace (grossly speaking a subset of wavelengths), which can be different at each stage. There is a specific subspace that discriminates the main classes; as shown later on in this paper, these classes correspond to the gross categories of blue and red galaxies. Then there are different latent subspaces to discriminate between the sub-classes within the main classes.

\section{Results}
\label{results}


In this section, we present the detailed full analysis of the subset of 302248 spectra that yield our proposed classification. The other subsets are discussed in Sect.~\ref{othersubsets}.

\subsection{Four main classes}
\label{fourmainclasses}

	   \begin{figure}
\begin{center}
		\includegraphics[width=\linewidth]{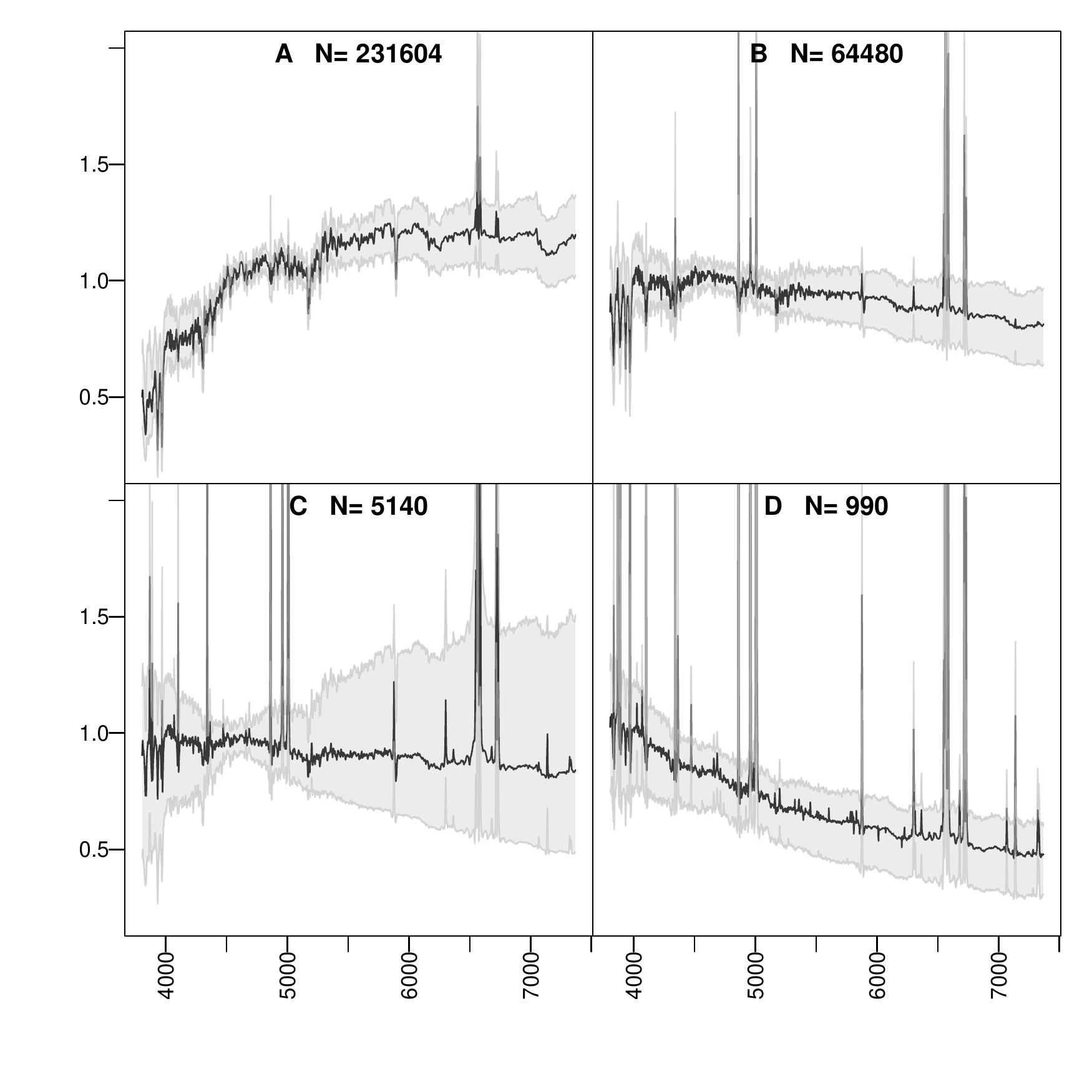}
	\caption{The four main classes for the subset of 301224 spectra. In all similar plots in this paper, the black line indicates the mean spectrum of each class, and the grey zone lies between the 10\% and 90\% quantiles for that class. The vertical scale is arbitrary and identical for all spectra. The class index and the number N of objects in the class are given in the legend. Classes are organised from left to right, top to bottom, in decreasing order of number of spectra.}
	\label{fig:threeclasses}
\end{center}
\end{figure}

A first analysis of the 302248 spectra with K=3 yields the following distribution: 301224, 990, 34 {(Fig.~\ref{fig:schemeanalyses})}. The smallest group is made of very peculiar spectra that we discard in this study, leaving 302214 spectra. 

The second {group (990 spectra) is not sub-clustered any further since its size is not too large and appears to be relatively homogeneous (Fig.~\ref{fig:threeclasses}): it gathers spectra with very blue continua and a lot of emission lines and is called class D in the rest of this paper.}

The largest group is then analysed with K=3 leading to the following distribution: 231604, 64480, 5140 (77\%, 21\% and 2\% respectively; {Fig.~\ref{fig:threeclasses}}). The two first classes (that we call A and B, respectively) are different in the slope of their continuum and the emission lines, and the third class C looks more diverse (Fig.~\ref{fig:threeclasses}).

{We end up with four main classes (A, B, C, and D).}
We now present the {sub-clustering analysis of first three of them (Fig.~\ref{fig:schemeanalyses})}. The sub-classes are named "Xn", where X is the main class (A, B or C) and n increases as the number of spectra within the sub-class decreases.

	   \begin{figure}
	   	\begin{center}
	\includegraphics[width=0.5\linewidth]{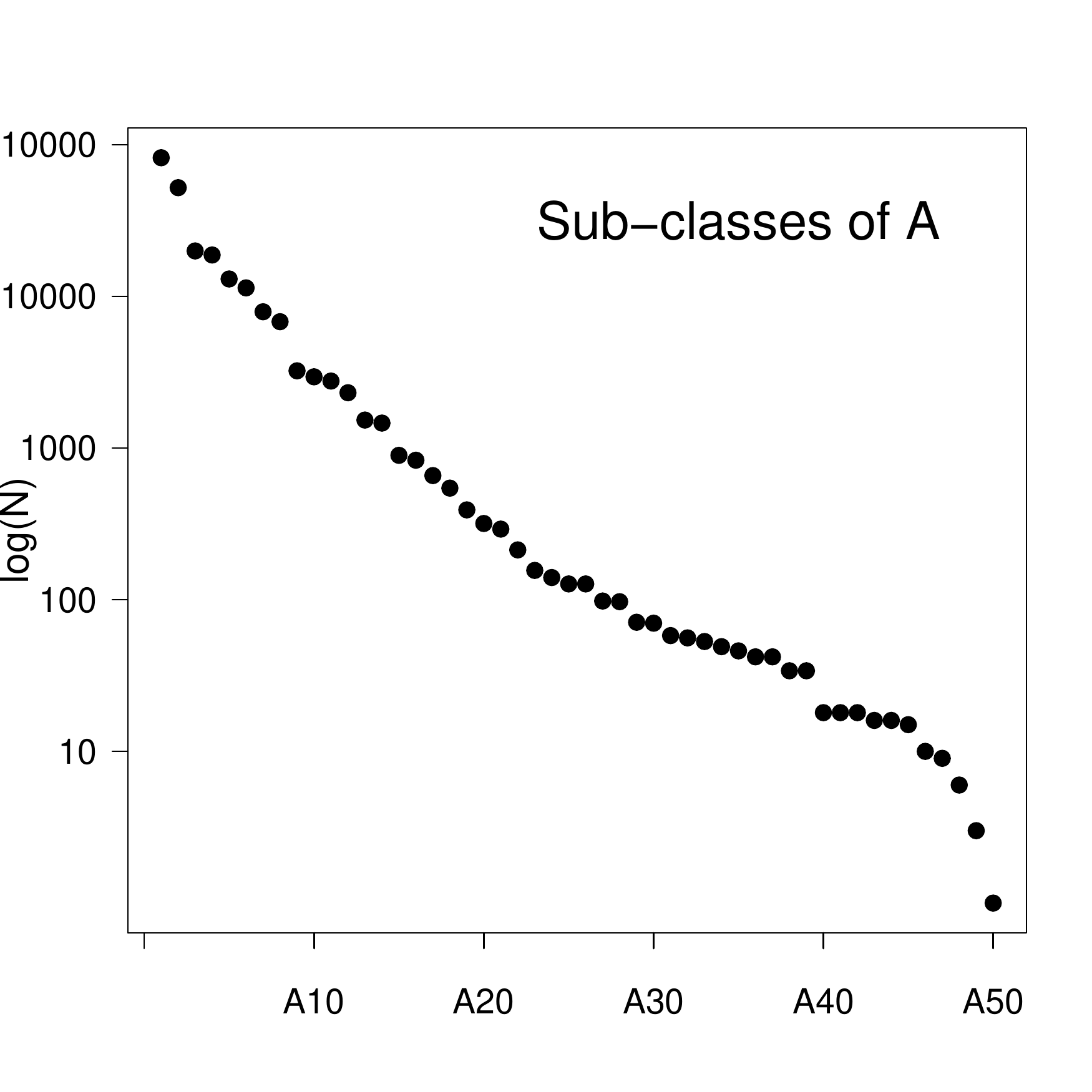}
	\includegraphics[width=0.5\linewidth]{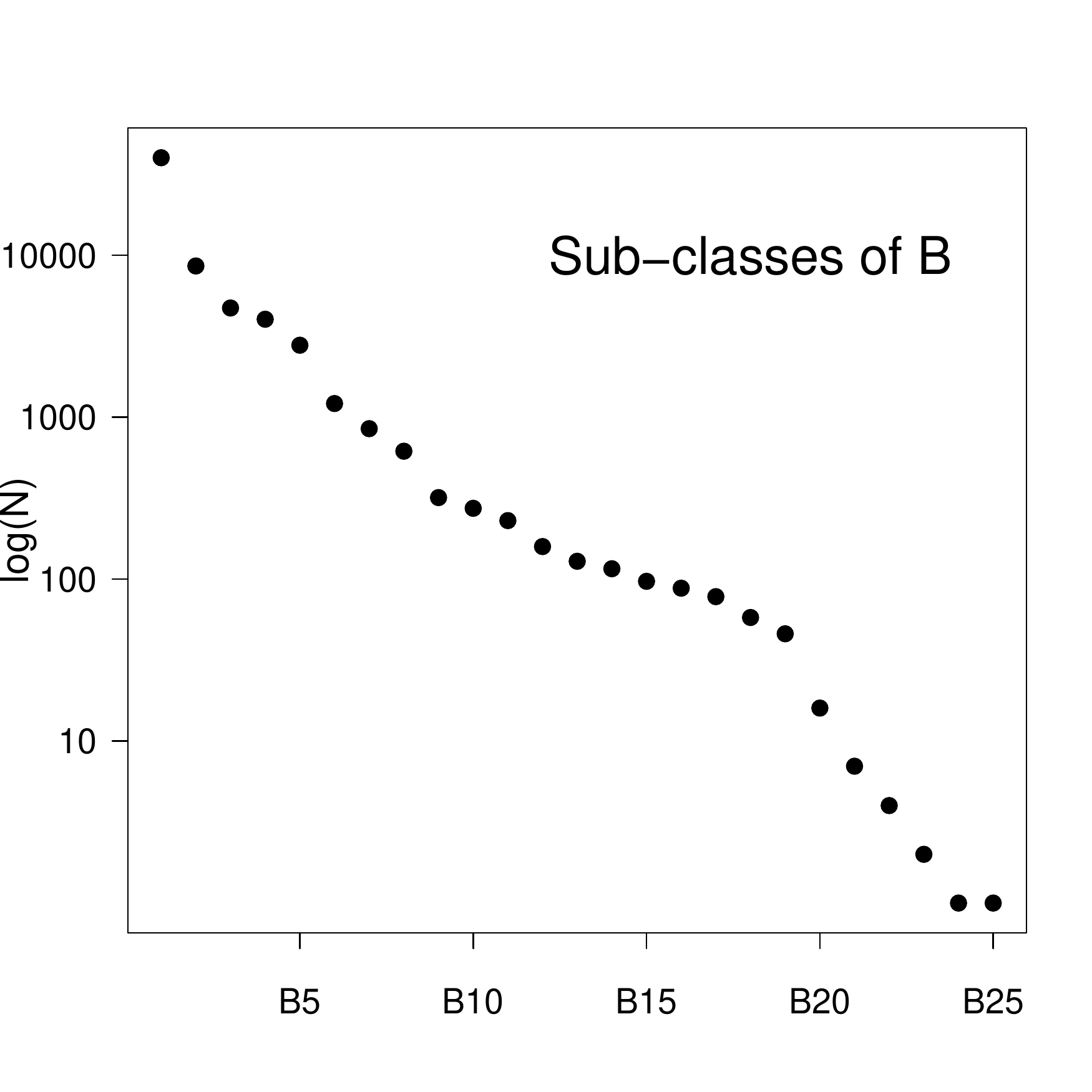}
	\includegraphics[width=0.5\linewidth]{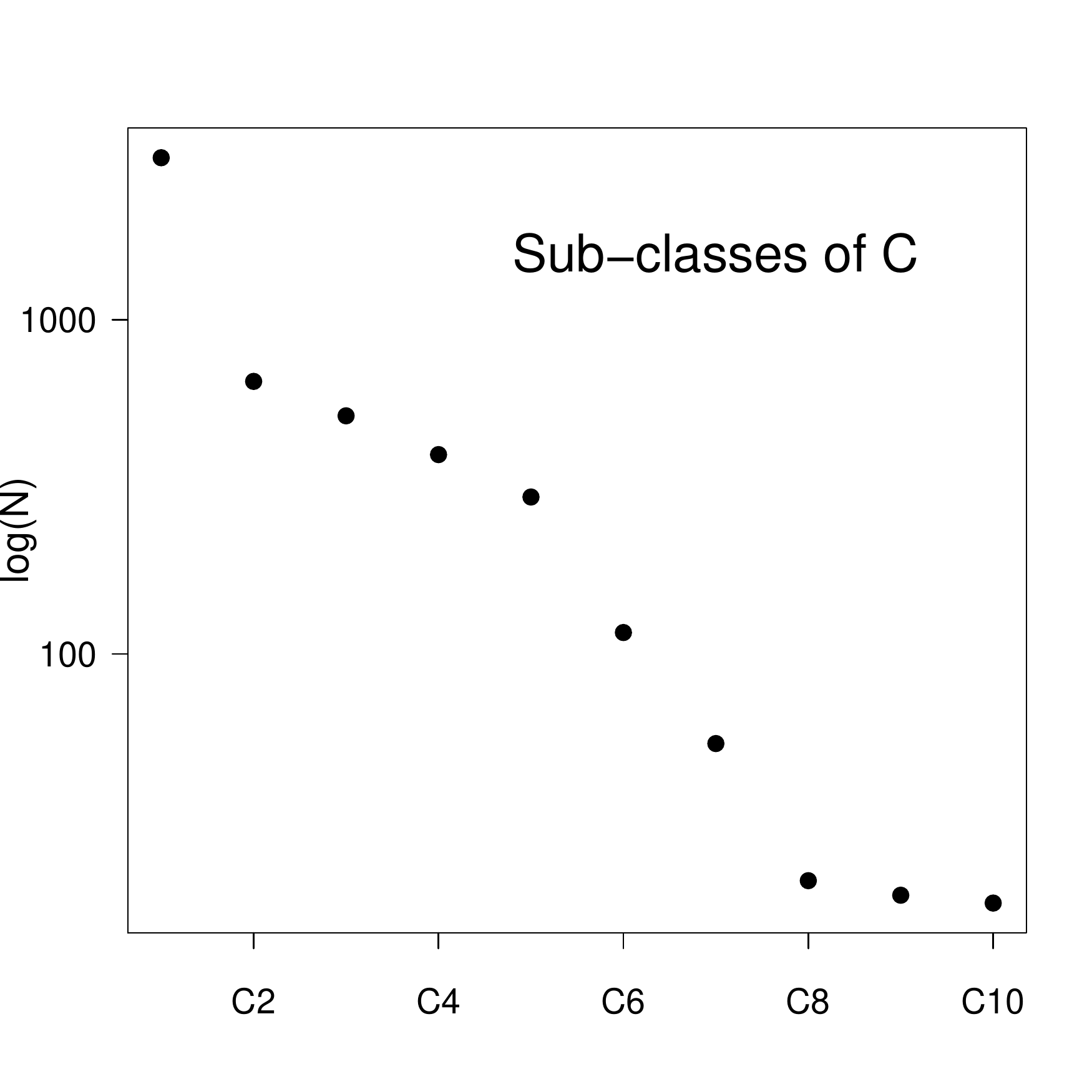}
	\caption{Distribution of {the number of} spectra in the sub-classifications.}
	\label{fig:ClassDistrib}
		\end{center}
\end{figure}

\subsection{Sub-clustering of class A}
\label{subclassA}

The curve of the ICL value (Sect.~\ref{fem}) as a function of K (Fig~\ref{fig:Aicl})) shows a strong increase and a significant inflexion at 50 before continuing with a flatter slope. Adopting a larger number of classes does not increase very much the ICL value, and brings many very small classes with less than 10 spectra. The optimum can thus be chosen at K=50. The distribution of {the number of } spectra among the sub-classes (Fig.~\ref{fig:ClassDistrib}) decreases nearly linearly in logarithmic scale, hence very sharply: the two most populated classes have 81992 and 52044 spectra (35\% and 22\%, respectively, of 231604), the other all being below 20000. 

The 50 sub-classes of class A are shown in Fig.~\ref{fig:ClassA}. The homogeneity (low dispersion) of the sub-classes is very good for the most populated ones, and tends to degrade for small sub-classes. We can already pinpoint very peculiar classes, such as A40 and A50 which probably contain very noisy spectra with some artefacts. Some others seem to contain strange features, such as A4 or A19, which have some strong absorption lines.

\subsection{Sub-clustering of class B}
\label{subclassB}

The ICL curve peaks at K=25 (Fig~\ref{fig:Bicl}). The distribution of spectra (Fig.~\ref{fig:ClassDistrib}) shows a dominant class (40040 spectra, 62\% of 64480), far larger than the others (8592, 4727, 4029, 2785, 1215...) with 5 very small classes of less than 7 members, two having only 1 spectrum.

The 25 sub-classes (Fig.~\ref{fig:ClassB}) are nearly all very homogeneous. Three sub-classes (B5, B9 and B11) show a redder continuum reminiscent of the class A, but they have strong emission lines that explain their presence here. The three classes B7, B8 and particularly B13, have wide emission lines.

\subsection{Sub-clustering of class C}
\label{subclassC}

The ICL curve shows an inflexion at around K=10 (Fig~\ref{fig:Cicl}) that we take as the optimum for the same reason as for class A above. For K=15, 3 classes have less than 4 members, so that we do not consider it a better classification. The distribution of spectra among the sub-classes (Fig.~\ref{fig:ClassDistrib}) shows a dominant class (3052 spectra, 59\% of 5140); the others have less than 700 members. This main class is very homogeneous like two similar but smaller classes (C6 and C7; Fig.~\ref{fig:ClassC}). These three classes differ in the detail of the emission and absorption lines, and they also are the only classes with a blue continuum. The other classes are more dispersed, and sometimes peculiar.

%
%

\subsection{Final classification}
\label{allclasses}

	   \begin{figure*}
\begin{center}
  \includegraphics[width=\linewidth]{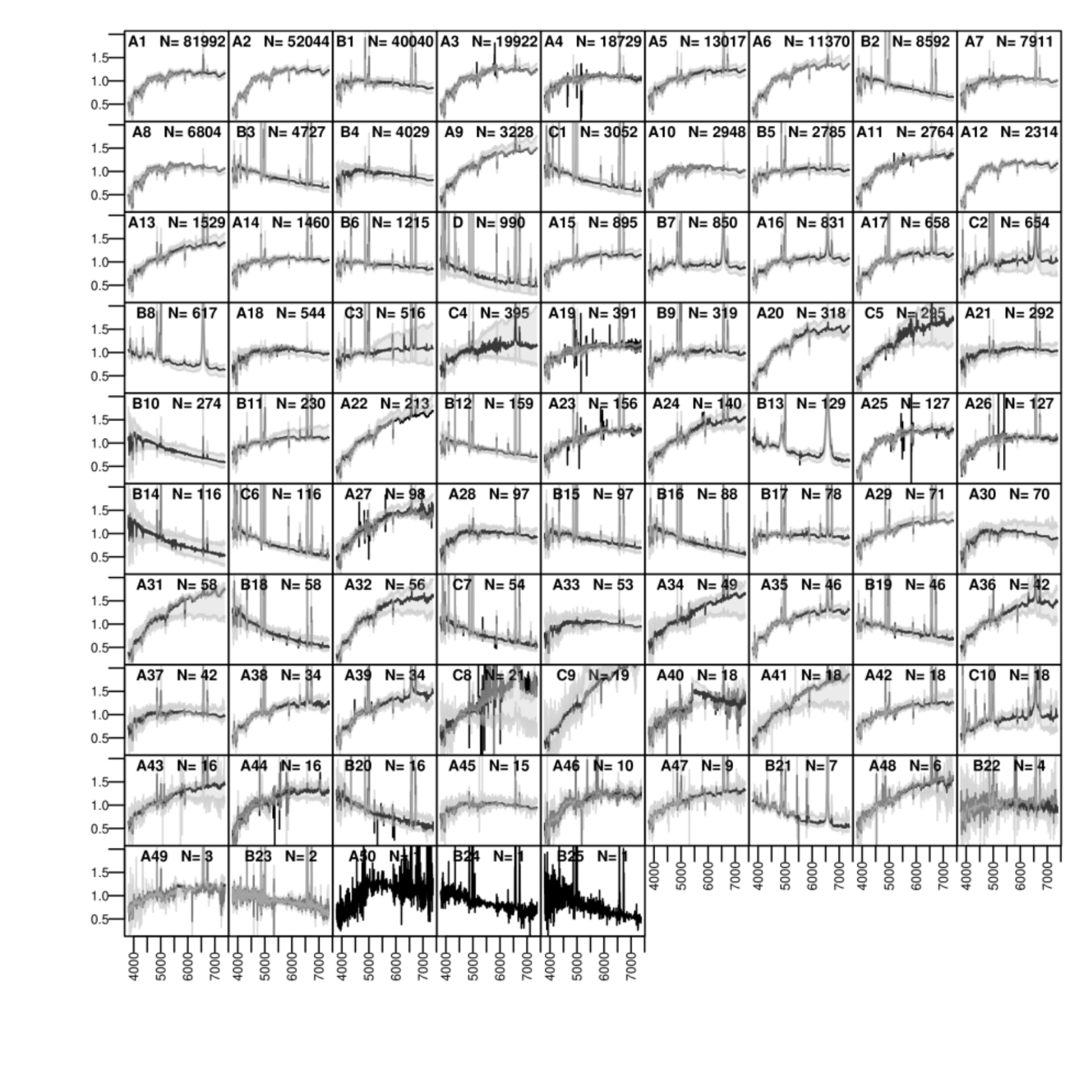}
  
  \caption{{Resulting 86 classes of the present study stored in decreasing order of the number of spectra within a class. The black line indicates the mean spectrum of each class, and the grey zone lies between the 10\% and 90\% quantiles for that class. The vertical scale is arbitrary and identical for all spectra. The class index and the number N of objects in the class are given in each graph.}}
	\label{fig:allclasses}
\end{center}
\end{figure*}

	   \begin{figure}
\begin{center}
		\includegraphics[width=0.7\linewidth]{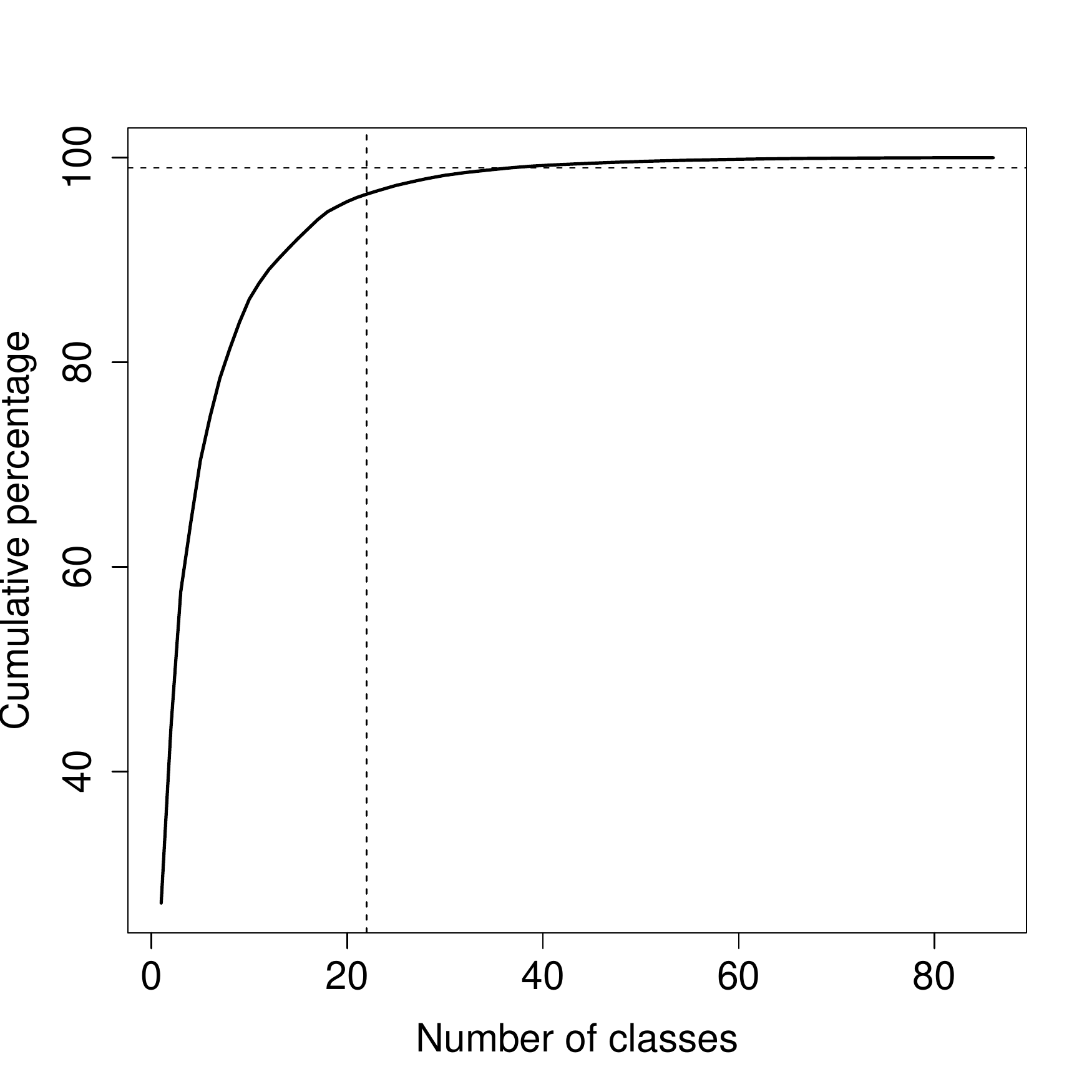}
	\caption{Cumulative percentage of spectra contained in the classes as shown in Fig.~\ref{fig:allclasses}. The vertical dotted line indicates the 22nd class (D) and the horizontal line the 99\% level crossing at 37 classes (see text).}
	\label{fig:cumulK86}
\end{center}
\end{figure}

{The sub-classes of the four main classes A, B, C, together with class D, all based on the 302248 subset of spectra, builds up our proposed classification in a scheme }
of 86 classes\footnote{The mean spectra of the 86 classes and the class of each of the 302248 spectra are available in electronic form
	at the CDS via anonymous ftp to cdsarc.u-strasbg.fr (130.79.128.5)	or via \url{http://cdsweb.u-strasbg.fr/cgi-bin/qcat?J/A+A/}} (Fig.~\ref{fig:allclasses}). They are organised according to their number of spectra. The largest class, A1, represents 27\% of the 302214 spectra, and the following classes 17\% (A2) and 13\% (B1), respectively. Class C1 (1\%) is at the 14th rank, and class D (0.3\%) at the 22nd. The 22 more populated classes are thus sufficient to cover the four main categories that we have identified and encompass 96\% of the spectra, while 99\% of the spectra are contained in the first 37 classes (up to class B9 which has 319 spectra, Fig.~\ref{fig:cumulK86}).

\section{{Analyses of the other subsets}}
\label{othersubsets}

{For comparison and robustness estimation, we also} performed complete analyses of the other set of 300000 spectra and of the 100000 subsets. For the full sample of 702248 spectra, we only performed the {pre- and first clustering, both} with K=3 to limit the computation load (see Sect.~\ref{choice}). We summarise the results in Fig.~\ref{fig:schemeanalyses}.  

The previous section shows that the first analysis with K=3 on the 302248 subset yields a very small group of peculiar spectra. It is remarkable that the full sample and that with 102248 (included in the 302248 spectra) reveal the same behaviour contrarily to all the other subsets. In other words there are some peculiar spectra contained and individualised in the 102248 sample that are also individualised in the larger samples that contain them. However, they cannot be the same since there are 7 in the 102248 subset, and 34 and 493 in the larger subsets. Even though they always represent a tiny fraction of the samples (less than 0.07\%), they would deserve further investigation, which we reserve for a subsequent paper.

One lesson of this finding is that there are very few spectra that are different from the vast majority. On one side this is unfortunate since they tend to disturb the clustering analysis imposing a process in several steps somehow depending on the subset. On another side, this demonstrates the discriminative power of the algorithm Fisher-EM which can be explained at least partly by the use of a latent subspace (Sects.~\ref{fem} and \ref{choice}).

For the three sets above (102248, 302248 and 702248), the second classes are considered as being class D as confirmed by the similarity of their spectra. This implies that this class D is absent from all the other subsets. 

The analyses with K=3 of the largest class of the three sets above and of all the other subsets (of 300000 and 100000 spectra) yield very similar distributions of spectra in the three classes that can easily be identified as the main classes A, B, and C.

The optimum number of classes within each of the main class is somewhat different from one subset to the other, but is remarkably consistent. We decided not to further analyse classes of less than about 1500 spectra, which explains why in Fig.~\ref{fig:schemeanalyses} the number of class is given as 1 (right column). This concerns class C only. There might be several reasons for the mismatch between the optimum number of {classes}, but we can point to two main explanations. A first one is because the diversity is not homogeneous between subsets of spectra (they were designed arbitrarily), hence the classes are all not expected to be represented. A second explanation may come from the probabilistic nature of the Fisher-EM classification leading to some fuzziness in the boundaries between classes. 

{We conclude for a very good consistency between all these analyses and, as shown is the next section, there is a very good agreement between all the obtained classifications. For that reason, investing a large computation time in a thorough analysis of the full sample with 702248 spectra would not bring much more information than the classification obtained on the subset of 302248 spectra as presented in Sect.~\ref{allclasses}. The procedure presented in this paper may serve as a guide for other unsupervised clustering of larger samples.}

\section{Robustness estimation}
\label{robustness}

In this section we assess the robustness of the proposed classification into 86 classes through a five-fold cross-validation technique and using the analyses of the different sub-samples presented in Fig.\ref{fig:schemeanalyses} (Sect.\ref{othersubsets}). 

For this purpose, we first present the two supervised techniques that we used.

\subsection{Supervised classification}
\label{supervised}

To assess the robustness of our classification, we need to match the classifications obtained with different samples, that is we need to classify a spectrum (typically the mean spectrum of a class) into the true classification obtained with the training sample. This is called supervised classification, and we used two techniques. 

The first technique uses the statistical model found by Fisher-EM with the training sample to compute the likelihood of the test spectrum on all classes running the E-step of the Fisher-EM algorithm. In other words, we compute the probability that the test spectrum belongs to each of the classes. This process is extremely fast and can be easily applied to classify millions of spectra.

The second technique is the k-nearest neighbour (kNN) algorithm. The (Euclidean) distance of the test spectrum with all the spectra of the training sample is computed and the class is determined from the majority vote of the k closest spectra. Obviously this method takes much more time than the previous approach. In addition, the Fisher-EM classification is obtained in a subspace of the features (wavelengths), while kNN uses all features to compute the distance between spectra. However, the kNN algorithm  is useful when there is no statistical model for the true classification. If the true classes are available only through their mean spectra, then k=1. When the class attribute is given for the whole training sample, then k=5 is a standard choice and is adopted in this study.

The error rate provided in the \R\ package \textit{mclust} is used in this paper as an indicator of the quality of the match between two classifications.  This rate gives the proportion of misclassified spectra in a minimum error mapping (correspondence) between the predicted classification and the true classes. 
We note that the comparison between our classifications is hampered by the relatively high number of classes (nearly 100) and their extreme imbalance; the classes have from one up to several tens of thousands of spectra).

Hereafter, the reference classification is that obtained with the reference sample of 302214 spectra and corresponds to the 86 classes shown in Fig.~\ref{fig:allclasses}. This should not be confused with the true classification that depends on the training sample.

\subsection{Cross-validation}
\label{crossvalidation}

We performed a five-fold cross-validation in the following way. The reference sample of 302214 spectra was split into five disjoint sets of about 60000 spectra. For each such set, we performed the full Fisher-EM analysis on the remaining 240000 spectra (the training sample) and used the resulting statistical model to predict the classes of the 60000 spectra (the test sample). We finally compared the predicted classes to the reference classification.

The optimum number of classes for the five cross-validation subsets of 240000 spectra is between 74 to 116 (as compared to 86 for the reference classification), and the error rate on the five 60000 sets is between 25\% to 42\% with an average of 34\%. This is called the test error.

We also performed some other tests. For instance, we estimated the training error by predicting the class of each spectrum of the reference sample (302214) using the statistical model of the reference classification. This error is found to be zero showing the quality of the classification despite the large number of small classes.

We also applied kNN with k=5 on 100 sets of 1000 randomly selected spectra out of 302214 and found a mean error rate of 11\%. We could not do the test on the full sample for excessive computation time.

\subsection{Comparison with the analyses of other subsets}
\label{compothersubsets}

\subsubsection{Distinct sets of 100000 spectra}
\label{other100m}

The optimum number of classes in the subsets of 100000 spectra is between 21 to 61 (Fig.~\ref{fig:schemeanalyses}), hence always lower than 86. This may indicate that the diversity of spectra in the training sample is not represented in all the 100000 subsets. To ensure a better match between the classifications, we first need to reduce (merge) the reference 86 classes to the number of classes for each subset. This is done with the function \textit{mapClass} from the package \textit{mclust} in \R or by matching the mean spectra of the classes with kNN (k=1). 

The error rate is found to be between 9\% and 21\% with a mean at 15\% for the six subsets of 100000 spectra and the subset with 102248 spectra.

\subsubsection{A distinct set of 300000 spectra}

\label{other300m}

The optimum total number of classes 86 as for the reference classification, but obviously the classes are not exactly the same since the class D is absent (Sect.~\ref{othersubsets}). By matching the number of classes as previously, the error rate is found to be about 30\%.

Even though this result is quite satisfactory because the reference and test samples are of similar sizes, it is intriguing that the error is higher than for the smaller samples with 100000 spectra. We find a clear correlation (Pearson coefficient of about 0.9) between the number of classes and the error rate. Our interpretation is that the culprit is probably the large number of small classes: half of these gather less than 1\% of the sample. {By construction, the ICL criterion (Sect.~\ref{fem}) avoids over-fitting as much as possible and favours well-separated clusters \citep{Biernacki2000}, but the nature of very small classes in the context of unsupervised clustering is not so clear. We wonder whether they are due to some kind of over-fitting, are real outliers or are peculiar objects heavily affected by artefacts. Unsupervised classification is a guide towards understanding the sample and only a detailed physical analysis of the spectra in the small classes can answer these questions.}

\subsection{Full sample of 702248 spectra}
\label{full702m}

Since we did not perform a sub-clustering of the four main classes A, B, C, D, we estimated the error rate between the common set of 302248 spectra only for these and found 2.2\%. This shows the strong significance of the four main classes {that are found in all subsets. We infer that the four main classes are four very general categories of the galaxy diversity, as is confirmed by their physical interpretation (Sect.~\ref{discussion}).}

{Since the error rate between our reference classification and the error rates obtained from any subset of the full sample is less than 30\%, and given the large size of our subsets and the large imbalance of the classifications, we conclude that the 86 classes obtained from the subset of 302248 spectra are fully representative of the full sample of 702248 spectra. }

{Our strategy to study several subsets of the full sample is successful in estimating the robustness of the proposed classification. This gives us confidence in using this approach for the analysis of other large samples. Instead of simply increasing the sample size to get stuck into computation issues, it is more efficient to study several smaller subsets to identify one that is representative of the full sample. In our case, it seems clear that the subset of 300000 spectra does not give more information than that with 302248, which uniquely contains the main class D. The important point is that the representativeness is not given by a priori hypothesis on what the diversity should be, but on purely clustering arguments.}

{Finally, robustness does not mean that every single class is exactly reproducible. We are working in a very high dimensional space, in which we are looking for the best optimum we can find, not the best in absolute terms if it exists. Smaller classes are probably less robust than the larger classes, but are not necessarily less interesting.}

\section{Discussion}
\label{discussion}

{  
	While the optimum number of classes is guided by an objective criterion (here the ICL value), there is still some arbitrariness in deciding how deep sub-clustering must be performed. We have seen that subdividing a class reveals more subtle categories, but it often produces very small classes with a handful of elements. The question arises as to whether this is useful. Naturally, this depends on the purpose of the study. There are fundamentally two goals: performing a supervised classification, that is deciding the nature of a new spectrum, or identifying peculiar objects. The algorithm Fisher-EM is able to tackle the two goals simultaneously by selecting categories during several steps and focussing on particular categories. In the following, we keep in mind both goals by ensuring that we keep peculiar classes in our study and that we drop out meaningless small classes.
}

\subsection{Description of the resulting classes}
\label{description}

{
We identified 86 classes, but a few of these can be discarded. For instance, the three smallest classes with only one spectrum (A50, B24 and B25) are clearly noisy (see Figs.\ref{fig:allclasses}, \ref{fig:ClassA} and \ref{fig:ClassB}). This also seems to be the case for classes with less than six spectra (A48, A49, B22 and B23). However, some very small classes look rather homogeneous such as B21 with seven spectra. It would certainly be useful to look at individual members of this class. 
On the other hand, the more populated class A40 can be discarded as well since it clearly gathers data with an artefact around 5000~\AA\ (Fig.\ref{fig:ClassA}). 
}

{
Apart from these noisy, inhomogeneous, and peculiar classes, most of the classes are very homogeneous (i.e. the mean, 10\% and 90\% quantile spectra are very similar), but in some cases, the intra-class dispersion is large as it is the case of most of the C sub-classes. This behaviour is especially important in the red part of the spectra, for example, in the case of A31 A32 A34, A36, A41 or B11 (see Figs.\ref{fig:ClassA} and \ref{fig:ClassB}).  
}

{
We should note that the dispersion around the mean spectra of the classes (Fig.\ref{fig:allclasses}) is always larger than about 10\%, hence much higher than the Galactic extinction we did not correct for (about 3\% in flux ratio of \textit{g} to \textit{r}). We cannot exclude some extreme cases that may impact our most dispersed and/or less populated classes which in any case will deserve further investigations in a subsequent study.
}

\subsection{Comparison with other atlases of spectra}
\label{comparison}

{
	In this section, we compare our classification with the atlases by \citet{Kennicutt1992}, \citet{Dobos2012} and \citet{Wang2018} described in the introduction (Sect.\ref{introduction}). The spectra of the comparison atlases were resampled to match ours and normalised to the mean between 4300 and 5000~\AA. The \citet{Kennicutt1992} spectra are shorter than ours, up to 6983~\AA\ instead of 7371~\AA, so that our spectra had to be shortened by 10\% in the comparison with this atlas. We must also recall that our spectra are not corrected for galactic dust extinction contrarily to the atlas spectra. But we here use the mean spectrum of each class, averaging out the extinction effect which is nevertheless small as compared to the intra-class dispersion (Sect.~\ref{description}).
}
	
	{
		Our classes are first matched with these atlases using the kNN (k=1) algorithm in Sect.~\ref{FEMtoatlas}. The reverse, that is the match of the atlas spectra with our classes, is done with the statistical model provided by Fisher-EM and presented in Sect.~\ref{atlastoFEM}. 
	}

\subsubsection{Match of our classes with the atlases}
\label{FEMtoatlas}

{
	The matches between each of our 86 classes with those of the atlases with the kNN (k=1) algorithm (Sect.~\ref{supervised}) are shown in Figs.~\ref{fig:FEMKen}, \ref{fig:FEMDobos} and \ref{fig:FEMWang}. The fit is generally good except for a number of spectra (see below). We must stress that since we have a much larger sample and more classes than these atlases, it is expected that many spectra and classes could not be present in the atlases, hence the fit is sometimes approximate.
}

{
	As can be noticed in the figures, our spectra are much better fitted by the \citet{Dobos2012} atlas (see Fig.\ref{fig:FEMDobos}) than the other two atlases (Figs.~\ref{fig:FEMKen} and \ref{fig:FEMWang}). For instance, the mean of the difference between the fitted spectra is always below 15\% (average 5\%) with an average standard deviation less than 20\% for the \citet{Dobos2012} atlas, except for our class B21 which is matched with a star-forming galaxy spectrum (SF1\_0).
	The agreement with the \citet{Kennicutt1992} atlas appears to be less good. The differences are always lower than 18\% (average 5\%) except for classes D and B21 because the standard deviations are on average 20\%.
For the \citet{Wang2018} atlas, the mean difference is lower than 31\% (average 7\%) and an average standard deviation of 20\%.

}

{
	This ascertainment is foreseeable since the \citet{Dobos2012} atlas is the most representative of the diversity of galaxy spectra comprising 32 classes and spanning three colours (red, blue and green) and six nuclear and star formation activities (from passive to seyfert galaxies, see their paper for more information). On the contrary, the two other atlases are either very poor in the number of classes or in the number of spectra representing a class. For instance, the \citet{Wang2018} atlas comprises six classes spanning nuclear and star formation activity and the \citet{Kennicutt1992} atlas is made of individual spectra (not averaged spectra) representing a single class. Hence the noise in the spectra of \citet{Kennicutt1992} atlas is higher and can disturb the kNN analysis. Moreover, the individual spectra in this atlas correspond to the classes defined only by the morphology of the galaxies.}
 
{
	In general, when the fit is not of good quality, the mismatch occurs in the red part of the spectrum where our spectra (in black) are often redder than those of the three atlases (shown in red).  The disagreement is most important in the fit of sub-classes of our class C, which is foreseeable since the dispersion in this class (and sub-classes) is very important; this can be seen in Fig.\ref{fig:ClassC} where the dispersion is represented by the spectra of the 10\% and 90\% quantiles. Also the mismatch in general may occur in the emission lines}, {probably ruling out the extinction correction difference between our spectra and those of the atlases}. {This behaviour seems to be more important in the fits with spectra from \citet{Wang2018} atlas. This is also foreseeable, since the spectrum representing a class in an atlas (except for the \citet{Kennicutt1992} atlas) is the average of the spectra in that class. As a matter of fact, the line intensities and ratios in averaged spectra are most probably not preserved. In addition, the kNN technique does not take the intra-class variance into account. In the case of the \citet{Kennicutt1992} atlas, a spectrum representing a class is an individual spectrum preserving the physical line ratios and intensities, but it corresponds to a given galaxy morphology that cannot be representative of the wealth of a galaxy class.
}

{
	The corresponding contingency table (Table~\ref{tab:corresp}) summarises the results of the fits of each of our classes with the three atlases. Our 86 classes are distributed into the 6 classes of \citet{Wang2018}, but in only 12 of the 24 classes of \citet{Kennicutt1992}, of which the missing classes are cD, cG, dI, E, IBm, Im, Sb:, SBa, SBb, Sbc, SBm and Sm. These are also distributed into 26 of the 38 classes of \citet{Dobos2012}, of which the missing classes are GG, h\_G, h\_GG, h\_RG, hh\_BG, l\_GG, l\_RG, p\_RG, RED0\_0, RED1\_0, RG and t\_BG (see Table~\ref{tab:corresp} for the meaning of their nomenclature). As mentioned before, in the \citet{Kennicutt1992} atlas, the categories are defined solely from morphology, not spectral features, while the \citet{Dobos2012} atlas has many overlapping categories.
}

\subsubsection{Match of the atlases with our classes}
\label{atlastoFEM}

{
	Conversely, with the use of the statistical model provided by Fisher-EM, we classified the spectra of the three atlases into our own classification. This is shown if Figs.~\ref{fig:KenFEM}, \ref{fig:DobosFEM} and \ref{fig:WangFEM}. We also performed a kNN (k=5) fit and the result is quasi-identical.
For the \citet{Dobos2012} and \citet{Wang2018} atlases, the agreement is pretty good, apart from some emission lines that are clearly not matched well. For the \citet{Dobos2012} atlas, the difference in the match has a mean always lower than 4\% with a standard deviation of less than 8\%, except for the class SF0\_0 for which these values are about 9\% and 13\%, respectively. For the \citet{Wang2018} atlas, these values are 4\% and 6\%, respectively.
Yet, our class A1 seems over-represented. 
}

{
	The match with the \citet{Kennicutt1992} atlas is less good, and is particularly bad for the spectra cG, I, and Sm, which are very noisy, and those of dI, Im, S, S0 (2 spectra), Sb:, and SBm, which are all best fitted with our D class or one sub-class of our C class. This result shows that our C class is mainly defined by strong emission lines. As a matter of fact, the spectra of the \citet{Kennicutt1992} atlas that are less well fitted are those with strong emission lines and could only be fitted (badly) with those of class C and D which are characterised by strong emission lines. Apart from these classes, the differences have a mean of less that 6\% and a standard deviation of 34\% at most.
}

		\begin{figure}
			\includegraphics[width=\linewidth]{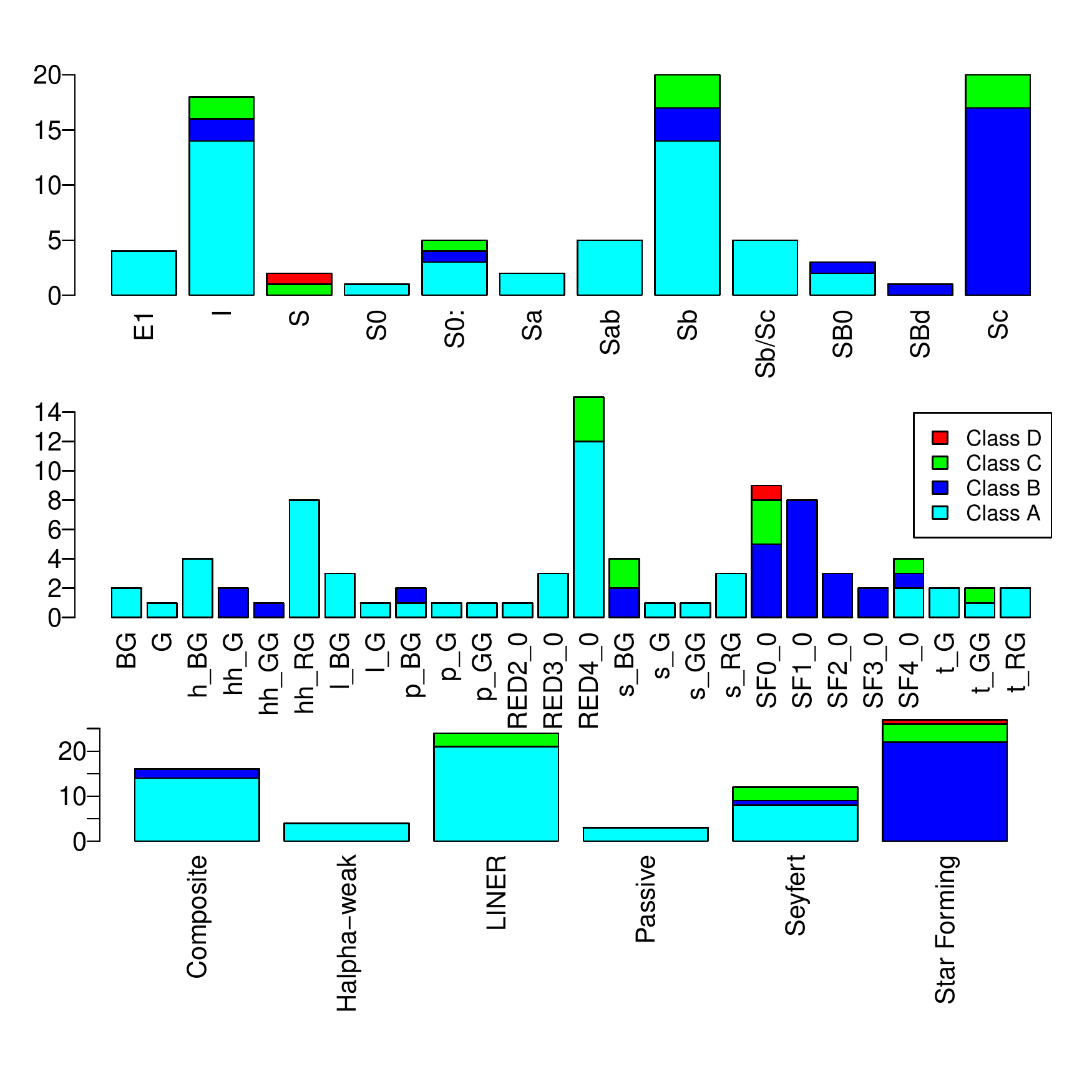}
			\caption{Distribution of our classes A, B, C and D among the classifications by \citet{Kennicutt1992}, \citet{Dobos2012} (see nomenclature explanation in Table~\ref{tab:corresp}) and \citet{Wang2018} from top to bottom.}
			\label{fig:FEMcontable}
		\end{figure}
                
{
	Figure~\ref{fig:FEMcontable} provides a summary of the distribution of the atlas categories with respect to our four main classes A, B, C and D. The class A is diverse in morphology while class B is mainly composed of Sc type galaxies according to the \citet{Kennicutt1992} atlas. Following the \citet{Dobos2012} atlas, our class B is clearly made of star-forming galaxies (plus AGN+HII), class A gathers many kinds of galaxies with little star formation, and class C is more difficult to characterise as with the \citet{Kennicutt1992} atlas. Finally, the simple \citet{Wang2018} atlas confirms that class B is made of star-forming galaxies such as class D, class C comprises LINERS, Seyferts and star-forming galaxies, and class A gathers everything except star-forming galaxies. 
}

\subsection{Physical interpretation of the classes}
\label{physinterpretation}

{
	From the above description of our classes and the comparison with other atlases of spectra, we can make an initial attempt to draw a physical meaning of our classes. A much deeper analysis will be made in a future work to provide a more precise interpretation.
}

{Class A is definitely composed of red galaxies, with a gradual activity of their nucleus or of star formation. The passive galaxies  are represented for example by classes A8, A10, A12, A18 with different reddening of the continuum, a slight activity is represented for example in classes A1, A2, A6, A9, A31, A32... with a gradual strength of the H$_\alpha$ line and an absence of H$_\alpha$. A number of sub-classes of class A are clearly representative of LINER galaxies with gradual values of the line ratios [OIII]/H$_\beta$ and [NII]/H$_\alpha$, this is the case for example of classes A3, A4, A5, A7, A11, A13, and A14. There are also transition objects at different quenching stages (A13, A14...).  Finally, in class A, we find also sub-classes that represent Seyfert 2 galaxies such as A16, A17, A19, and A23.
}

{
	Class B is made of bluer galaxies with a higher activity deduced from the stronger and higher number of emission lines. This class is representative of different levels of star-forming galaxies (example classes B1, B2, B3, B4, B6 ...) but also seemingly, of Seyfert 1 type galaxies as suggested by the spectra of classes B7, B8, B13 and B21. Finally some sub-classes show redder continuum (but bluer that class A galaxies) with a higher activity than class A. This is the case of classes B5 and B9.
}

{
	Classes C and D are characterised by strong and more numerous emission lines with the highest activity being in class D (see Fig.\ref{fig:allclasses}).
Indeed, emission lines dominate the spectra of class D and are all extremely blue. Consequently, Class D is definitely representative of high activity of star-forming galaxies.
Class C is defined exclusively by its set of emission lines that represent a transition class between class B and D in terms of activity. Hence, the spectra of class C span a wide range of colours from very red to very blue galaxies.
In Fig.\ref{fig:ClassC}, the sub-classes of class C confirm that this class is definitely characterised by the mission lines. As a matter of fact, the width of H$_\alpha$ line seems to define the sub-classes, for example classes C8 and C9 span wide ranges of continua but present a large H$_\alpha$ width. This is also the case for classes C2, C4 and C10 showing a single line width each that is different from one class to the other but a large variation in the continuum within a single class. The fact that our method is able to distinguish classes from the line width of their lines (for instance in our case, the H$_\alpha$ line) is a very promising and powerful result.
Finally, Class C is probably representative of galaxies with both star formation and nuclear activity, but a more thorough analysis of the spectra is needed to interpret its sub-classes.
}

\subsection{An illustration of the diversity of our 86 classes}
\label{scatterplots}

\begin{figure*}
	\includegraphics[width=\linewidth]{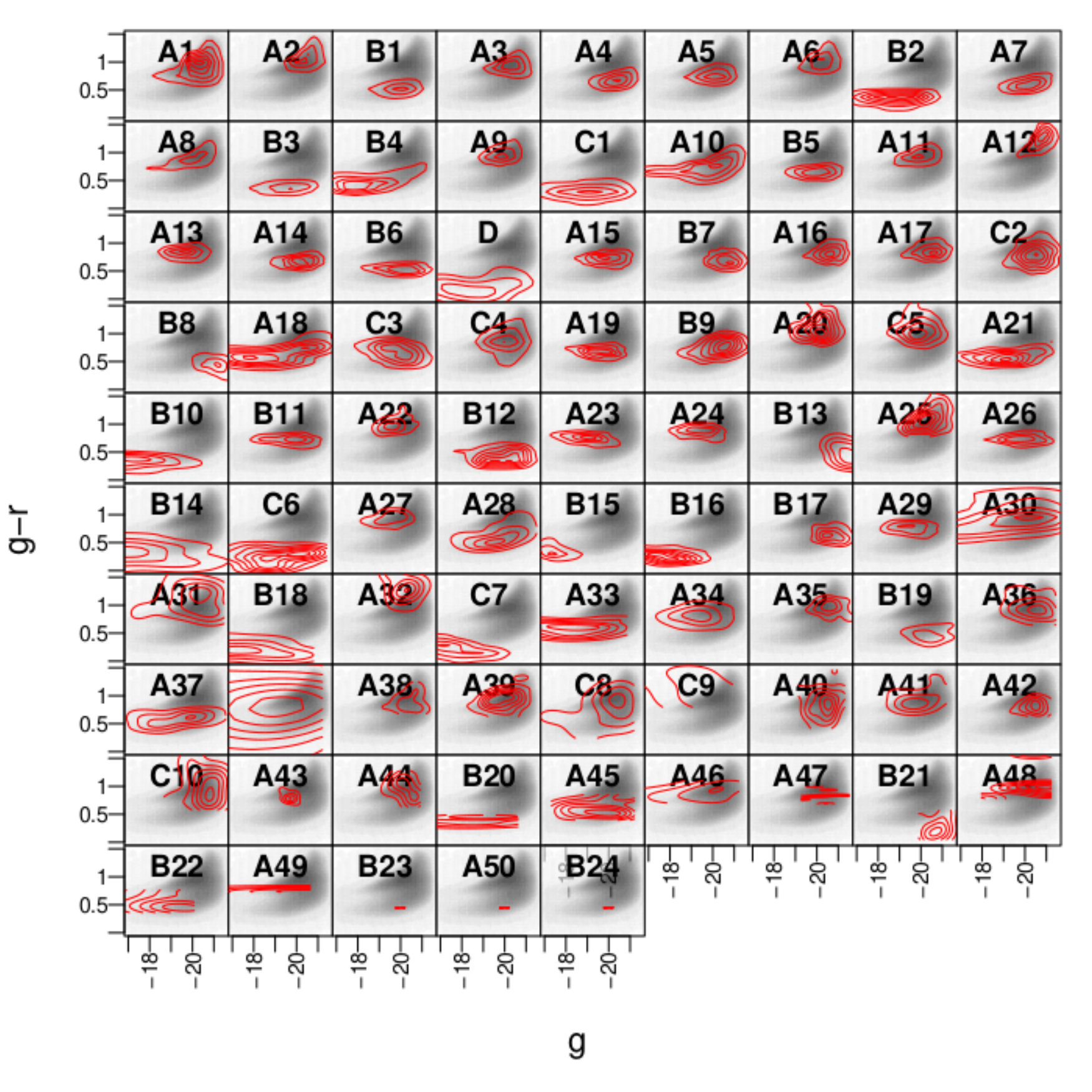}
	\caption{{Distribution of our 86 classes in the \textit{g-r} vs \textit{g} diagram. The distribution of all the 302248 objects used for the classification is shown in grey. The red contours show the distribution of the objects within each class.}}
	\label{fig:CMD}
\end{figure*}

\begin{figure*}
	\includegraphics[width=\linewidth]{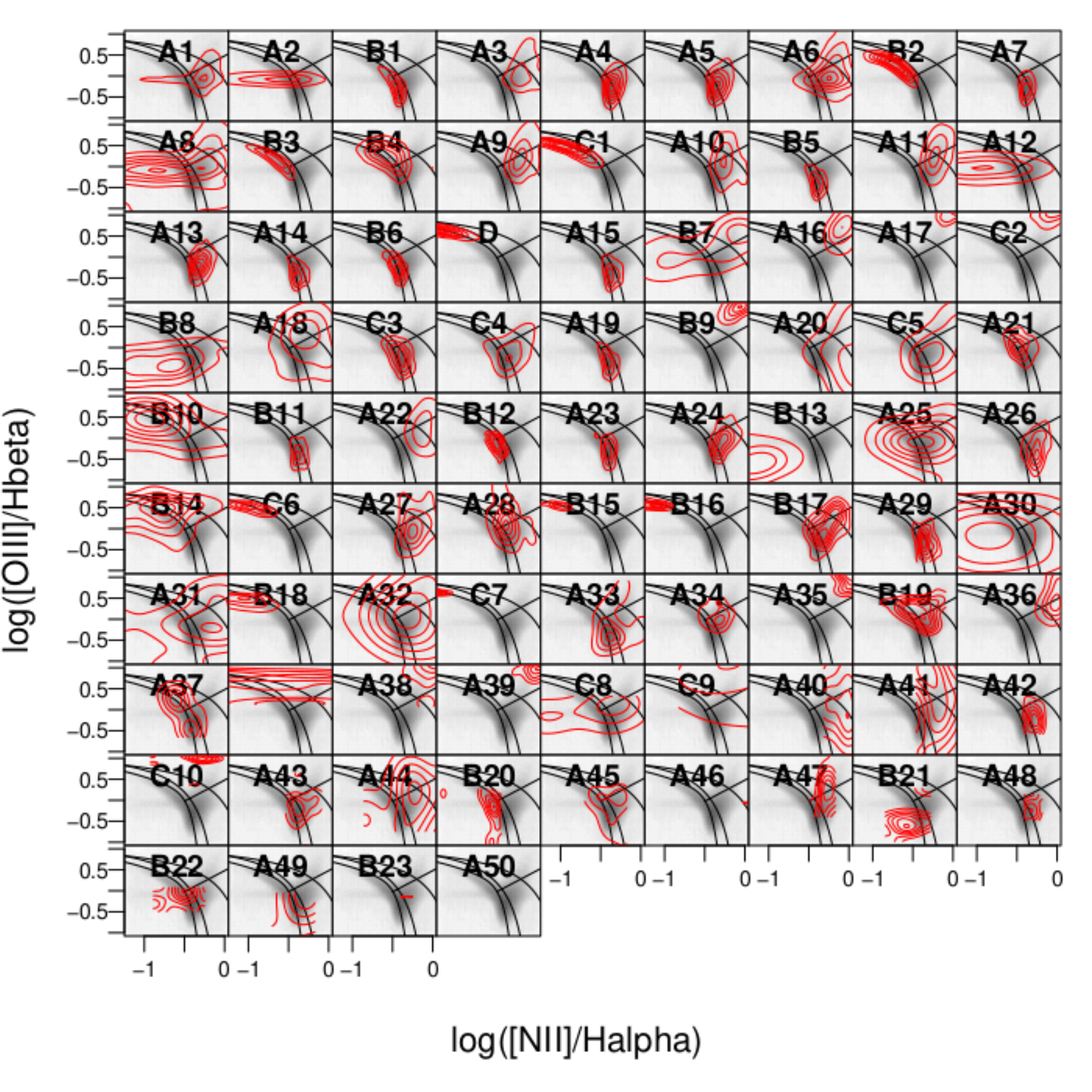}
	\caption{{Distribution of our 86 classes in the (BPT) diagram. The distribution of all galaxies with indices present in the DR7 release (not all the 302248 objects used for the classification) is shown in grey. The red contours show the distribution of the objects within each class. The curves are taken from \citet{Kewley2001,Kauffmann2003,Stasinska2006} (right to left respectively). The bottom left zone represents star-forming regions, the upper zone represents AGNs and Seyfert galaxies, and the bottom right zone represents LINERs.}}
	\label{fig:BPT}
\end{figure*}

{
	A detailed analysis of the properties of the classes is kept for a subsequent paper, but we wish to present some hints about the distribution of our classes in two famous scatter plots: a colour-magnitude diagram and the BPT \citep{Baldwin1981} which is often used to discriminate between star-forming and AGN-dominated galaxies, that is to determine the ionisation origin of the emission lines.
}

{
The colour-magnitude diagram (Fig.~\ref{fig:CMD}) confirms the analysis above based on the spectra themselves: the A class mainly belongs to the red branch, the B class to the blue branch. There are only very few exceptions to this scheme, such as A7 (blue branch) and may be B11 (blue branch). The C class is more diverse while the D class is obviously very blue and faint. This confirms that the main classes A, B, and D are discriminated by the slope of the spectra, whereas the main class C possess some other specific characteristics.
}

{
The BPT diagram of [OIII]/H$_\beta$ versus [NII]/H$_\alpha$ (Fig.~\ref{fig:BPT}) shows that A classes are generally situated around [OIII]/H$_\beta \simeq 1$, principally as star-forming or LINERs galaxies. 
The B classes are mostly on the verge between star-forming and other types, confirming that most of these galaxies are probably subject to intense star formation activity. The C classes kind of avoid the star-forming region (lower left zone) but are rather situated in the Seyfert, AGN of LINERs regions. Class D seems very specific, occupying a tiny region along the top left part of the star-forming limit, similarly to B15 or C7 for instance.
}

{
These two diagrams demonstrate that our classification obtained with the unsupervised clustering technique called Fisher-EM retrieves the diversity of the physical properties of galaxies with many details. They can usefully be compared with the classification in ten classes obtained by \citet{Chattopadhyay2019} using a very different set of features (photometry, spectroscopy, morphology, chemical composition, and kinematics) and a very different technique (ICA + k-means). Clearly, working with spectra is more direct and objective since the features do not have to be carefully selected, and yield a much more detailed classification.
}

\section{Conclusions}
\label{conclusions}

Using a latent subspace clustering algorithm called Fisher-EM, we were able to establish a robust and thorough classification of a large sample of spectra of galaxies from the SDSS DR7. The classification has been obtained with 302248 spectra, in a hierarchical process of sub-clustering the main classes. Cross-validation with the analyses of several subsets of our full sample (702248 spectra) demonstrates the robustness of our proposed classification.

The first step of classification with few classes always yields four main categories, grossly corresponding to quiescent, star-forming, Seyferts and LINERs, and galaxies with many emission lines. Subsequent clustering of these main classes yields a total of 86 classes, among which many very small classes are present, sometimes made of peculiar spectra, but also probably revealing possible interesting objects. 

In this paper, we did not attempt a detailed modelling of the mean spectra of the classes to derive the physical parameters that could characterise them. This is the purpose of a subsequent work. Instead, we matched our classification with three atlases in the literature by using either the statistical model provided by the Fisher-EM analysis or the simple kNN algorithm. The generally good outcome of the correspondence shows that an entirely automatic approach to build template spectra of galaxies is achievable and providing a way to perform a supervised classification on millions of spectra.

We more or less reach the same conclusion as \citet{Wang2018} when they compare their classification with that by \citet{Dobos2012} since the detailed match for emission lines is sometimes disappointing. Classifying complex spectra is certainly difficult, as illustrated for instance in \citet{Rahmani2018} in which a semi-supervised algorithm (self-organizing maps) is used to classify spectra against templates with a very small sample. However, we identify two possible caveats related to the use of templates. Firstly averaging spectra may lead to a loss of some precise and crucial information in the lines, thus we think this point should be addressed specifically on large samples. We intend to pursue the present work in this direction. In particular, we are completing two studies on simulated spectra of galaxies, one with simple single stellar populations (SSPs) and one using the CIGALE code \citep{Boquien2019}. Secondly, we believe it is important to take the intra-class variance into account when comparing with templates or other classifications. This is done automatically when using the statistical model provided by the Fisher-EM algorithm or any other kind of probabilistic clustering technique (e.g. GMM algorithms).

\begin{acknowledgements}
{We thank the referee for constructive comments that significantly improved the paper.} This work has been partially supported by the French government, through the 3IA Côte d’Azur Investment in the Future project managed by the National Research Agency (ANR) with the reference number ANR-19-P3IA-0002. {Funding for the Sloan Digital Sky Survey (SDSS) has been provided by the Alfred P. Sloan Foundation, the Participating Institutions, the National Aeronautics and Space Administration, the National Science Foundation, the U.S. Department of Energy, the Japanese Monbukagakusho, and the Max Planck Society. The SDSS Web site is http://www.sdss.org/.
The SDSS is managed by the Astrophysical Research Consortium (ARC) for the Participating Institutions. The Participating Institutions are The University of Chicago, Fermilab, the Institute for Advanced Study, the Japan Participation Group, The Johns Hopkins University, Los Alamos National Laboratory, the Max-Planck-Institute for Astronomy (MPIA), the Max-Planck-Institute for Astrophysics (MPA), New Mexico State University, University of Pittsburgh, Princeton University, the United States Naval Observatory, and the University of Washington. }
\end{acknowledgements}

%
   \bibliographystyle{aa} 
    \bibliography{FEMSDSS.bib}
%

\begin{appendix}
\section{Noise reduction and binning of original spectra}
\label{noise}

\begin{figure}[h]
\includegraphics[width=\linewidth]{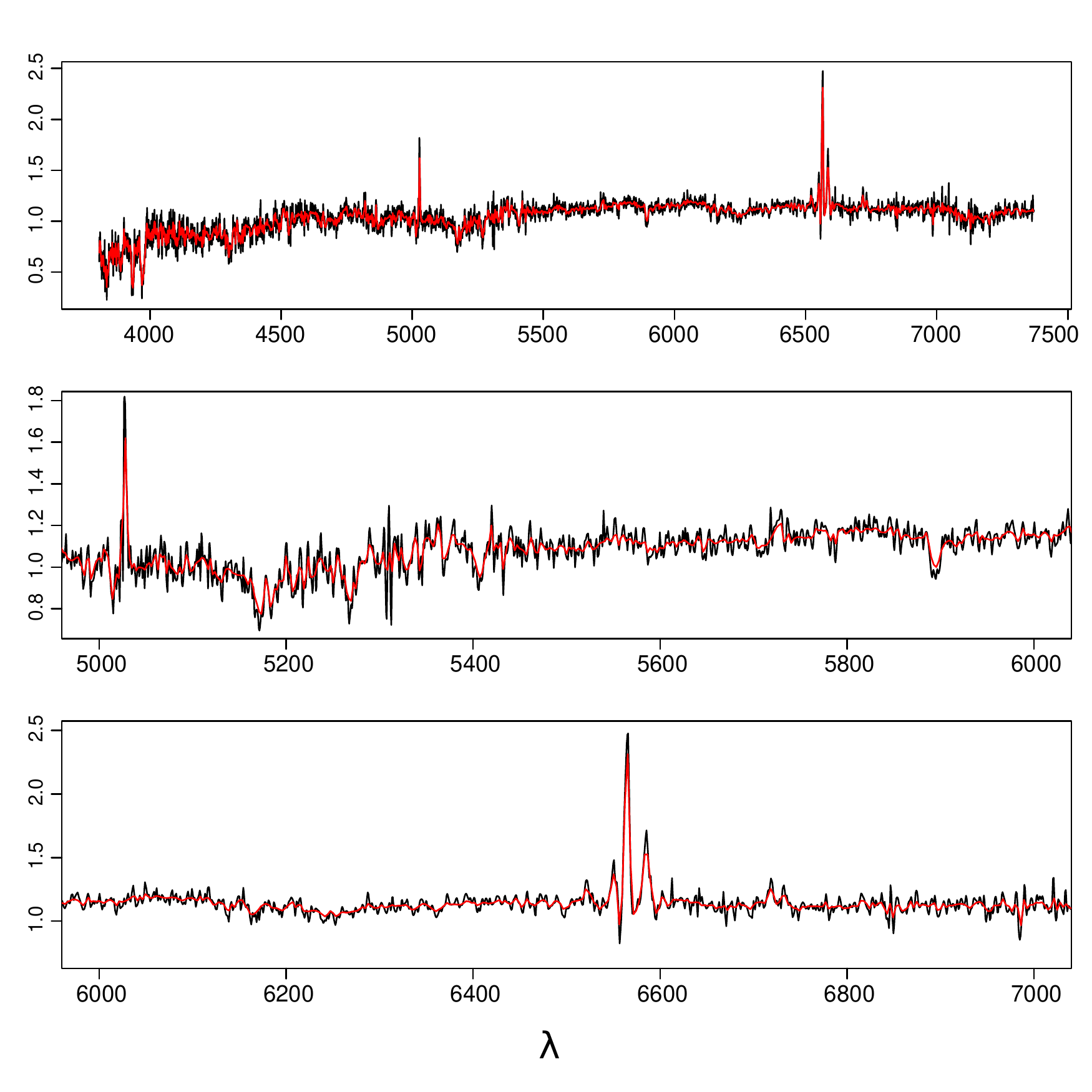}
\caption{One original spectrum (black) compared to the result after denoising and binning (red). \textit{Top:} Full range of wavelengths. \textit{Middle:} From 5000 to 6000~\AA. \textit{Bottom:} From 6000 to 7000~\AA.}
\label{fig:binningcheck}
\end{figure}

To preserve the shapes of emission and absorption lines as much as possible, each spectra was first resampled, increasing its size by a factor of two. We then applied a wavelet decomposition before reducing the noise by applying a threshold, that is by cutting the two shortest components that most probably carry the noise \citep{Donoho1994,SouzaFeliciano2018}. This step required some manual adjustments to set the best threshold levels by comparing some emission lines before and after noise reduction (Fig.\ref{fig:binningcheck}). Lastly, each spectrum was rebinned to reduce its length by a factor of 4, yielding 1437 wavelengths (see Sect.~\ref{data}).

\section{Sub-clustering}

\begin{figure}[h]
\begin{centering}	
\includegraphics[width=0.8\linewidth]{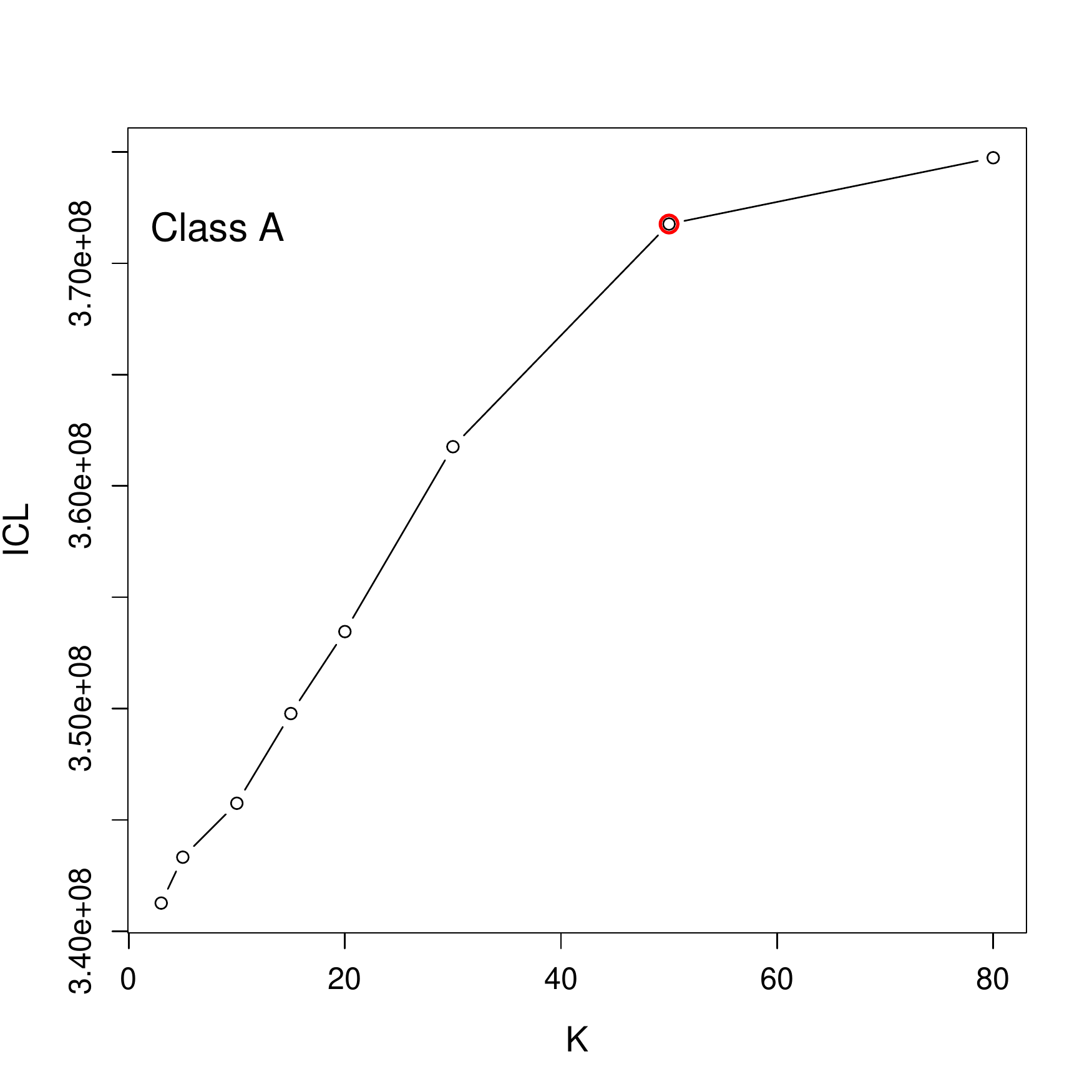} \\
\caption{ICL for different K for each class for the main class A. The red dot indicates the choice for K.}
\label{fig:Aicl}
\end{centering}
\end{figure}

\begin{figure}[h]
\begin{centering}	
\includegraphics[width=0.8\linewidth]{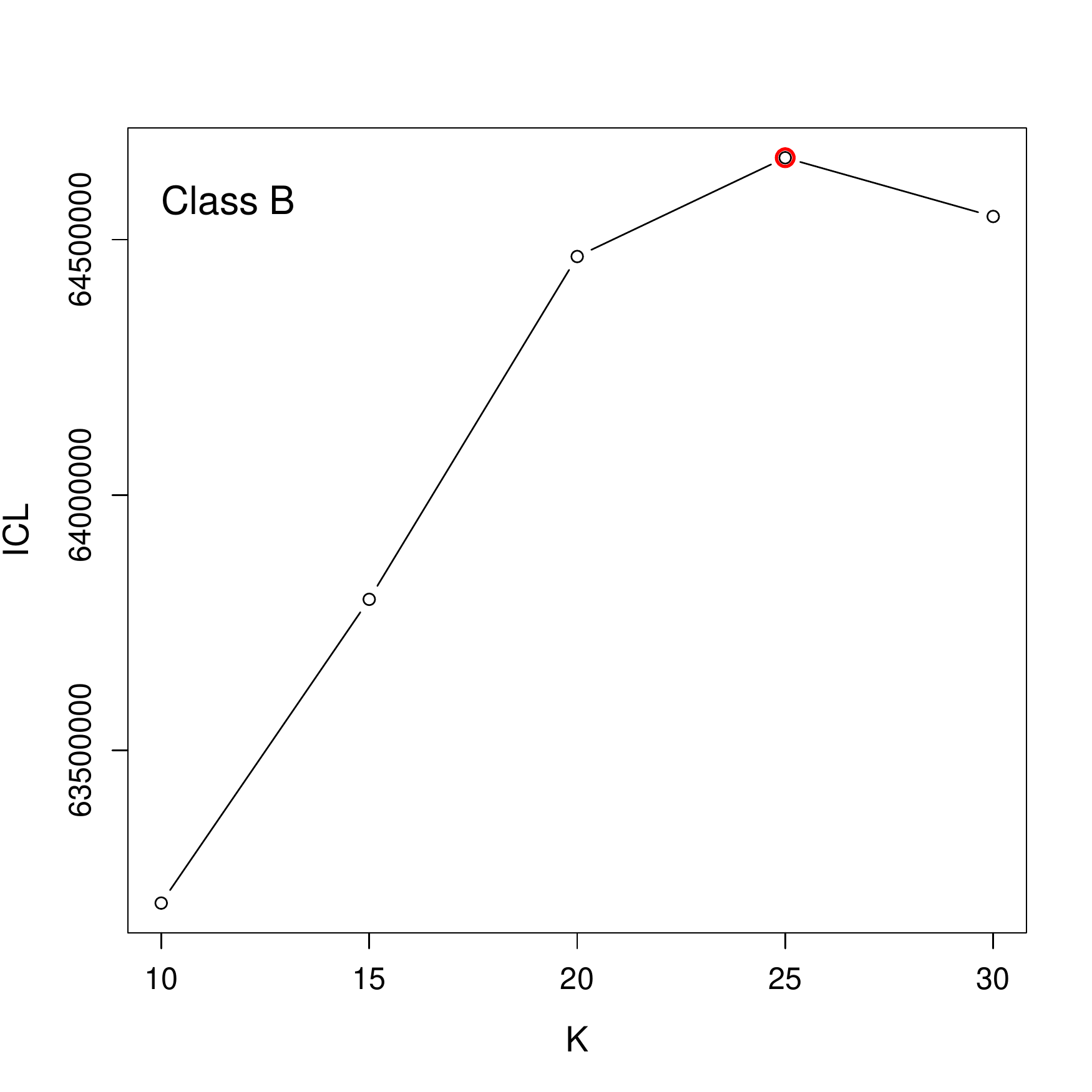} \\
\caption{Same as Fig~\ref{fig:Aicl} for main class B.}
\label{fig:Bicl}
\end{centering}
\end{figure}

\begin{figure}[h]
\begin{centering}	
\includegraphics[width=0.8\linewidth]{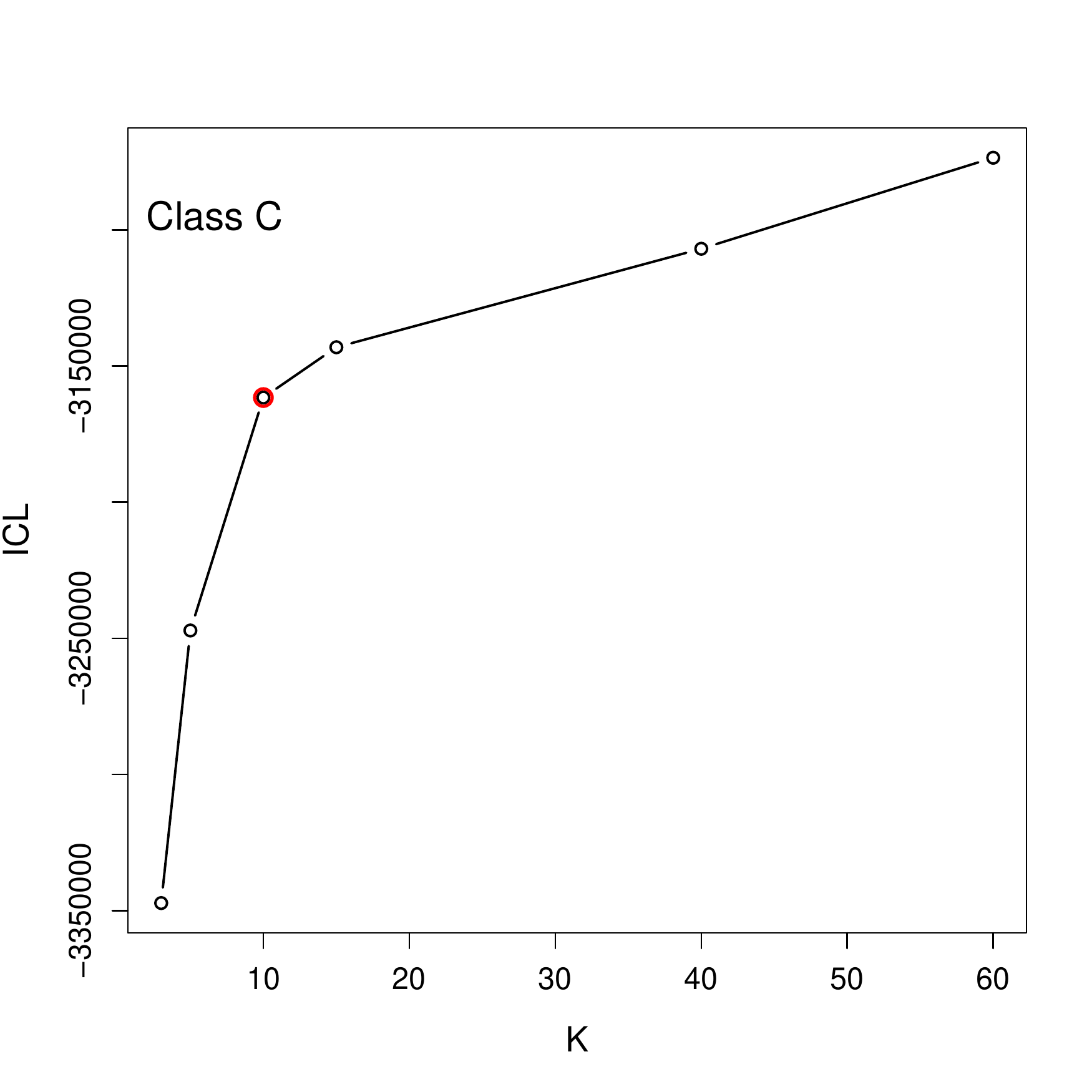}
\caption{Same as Fig~\ref{fig:Aicl} for main class C.}
\label{fig:Cicl}
\end{centering}
\end{figure}

\begin{figure*}
\begin{center}	
\includegraphics[width=\linewidth]{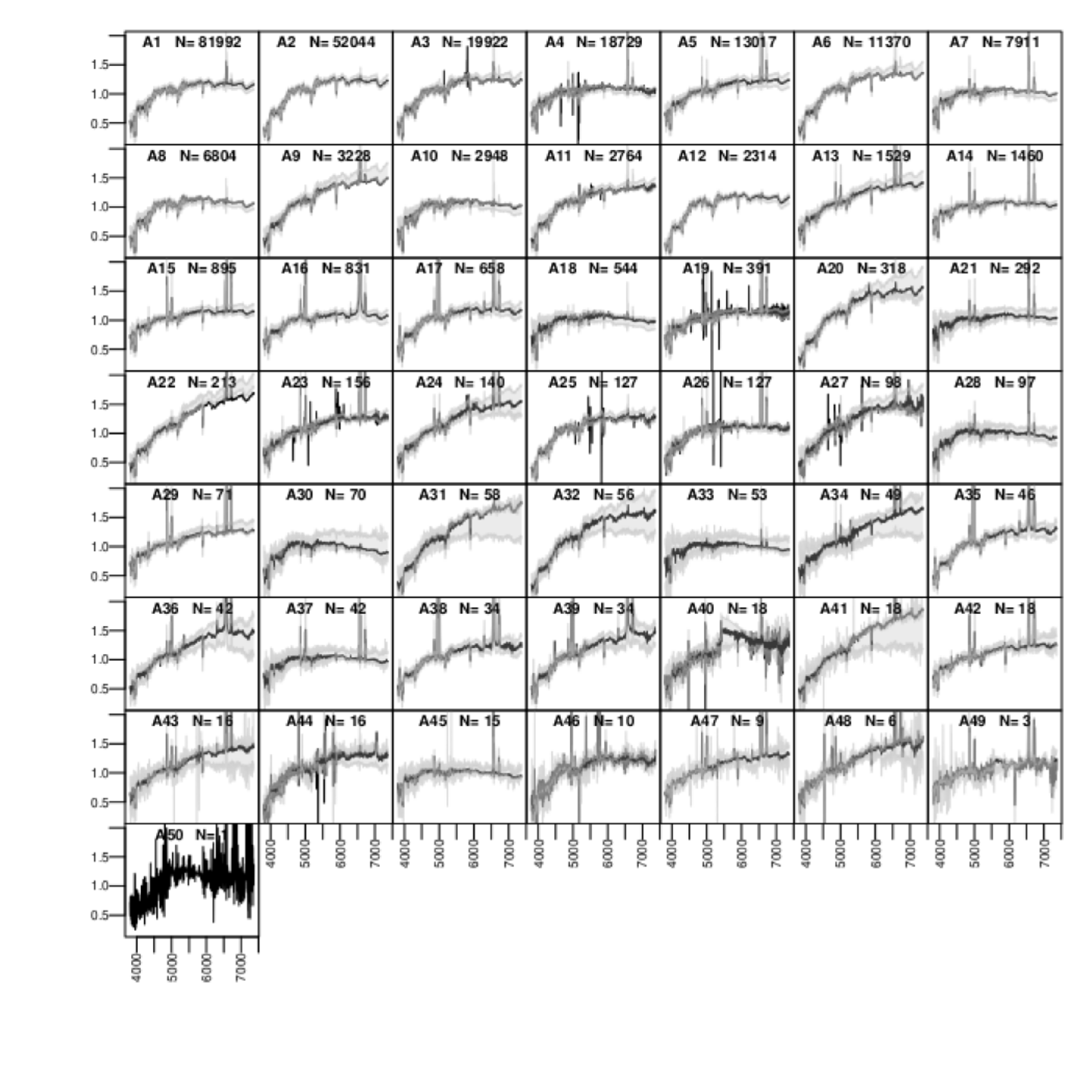}
\caption{Sub-classification of class A with 231604 spectra.}
\label{fig:ClassA}
\end{center}
\end{figure*}

\begin{figure*}
\begin{center}
\includegraphics[width=\linewidth]{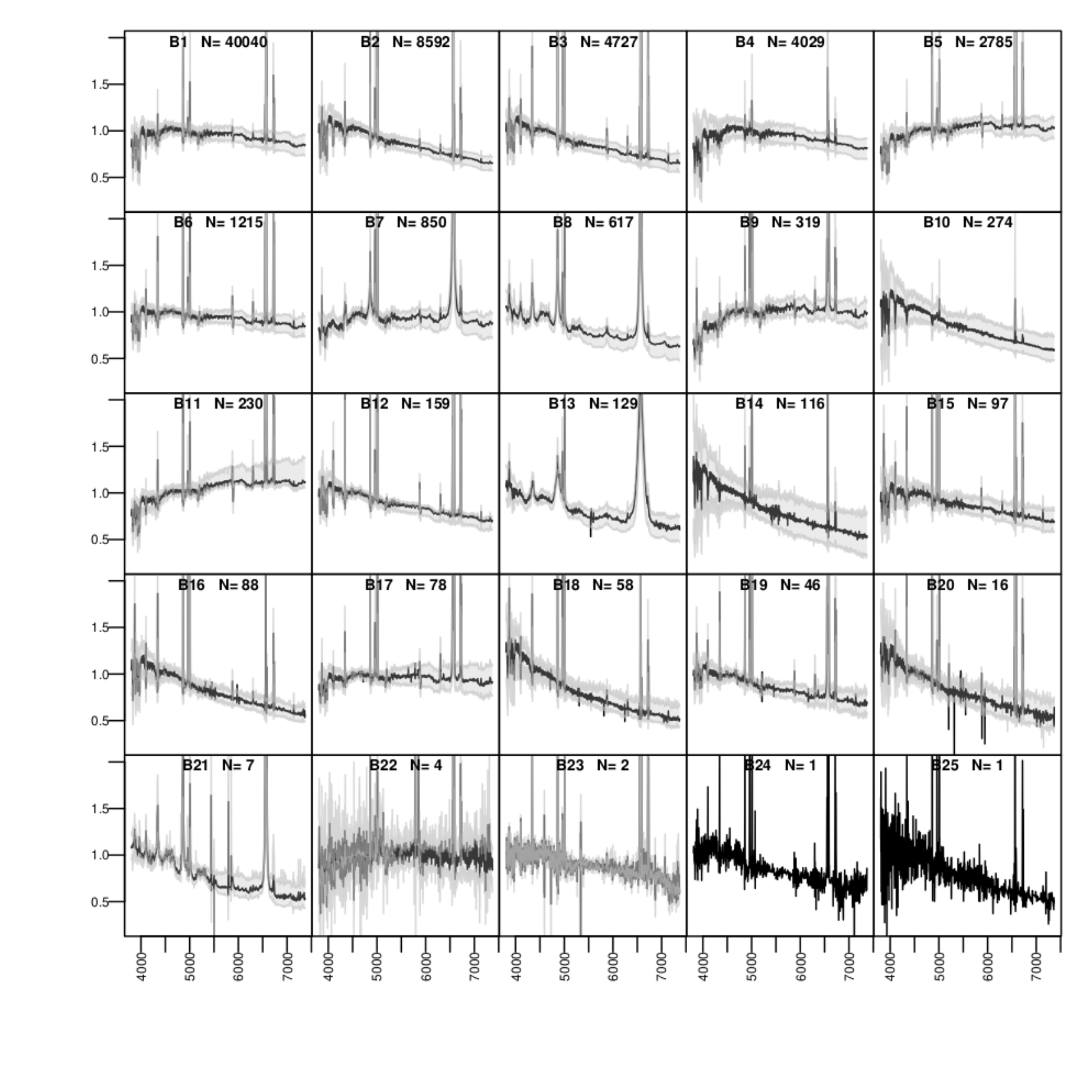}
\caption{Sub-classification of class B with 64480 spectra.}
\label{fig:ClassB}
\end{center}
\end{figure*}

\begin{figure*}
\begin{center}
\includegraphics[width=\linewidth]{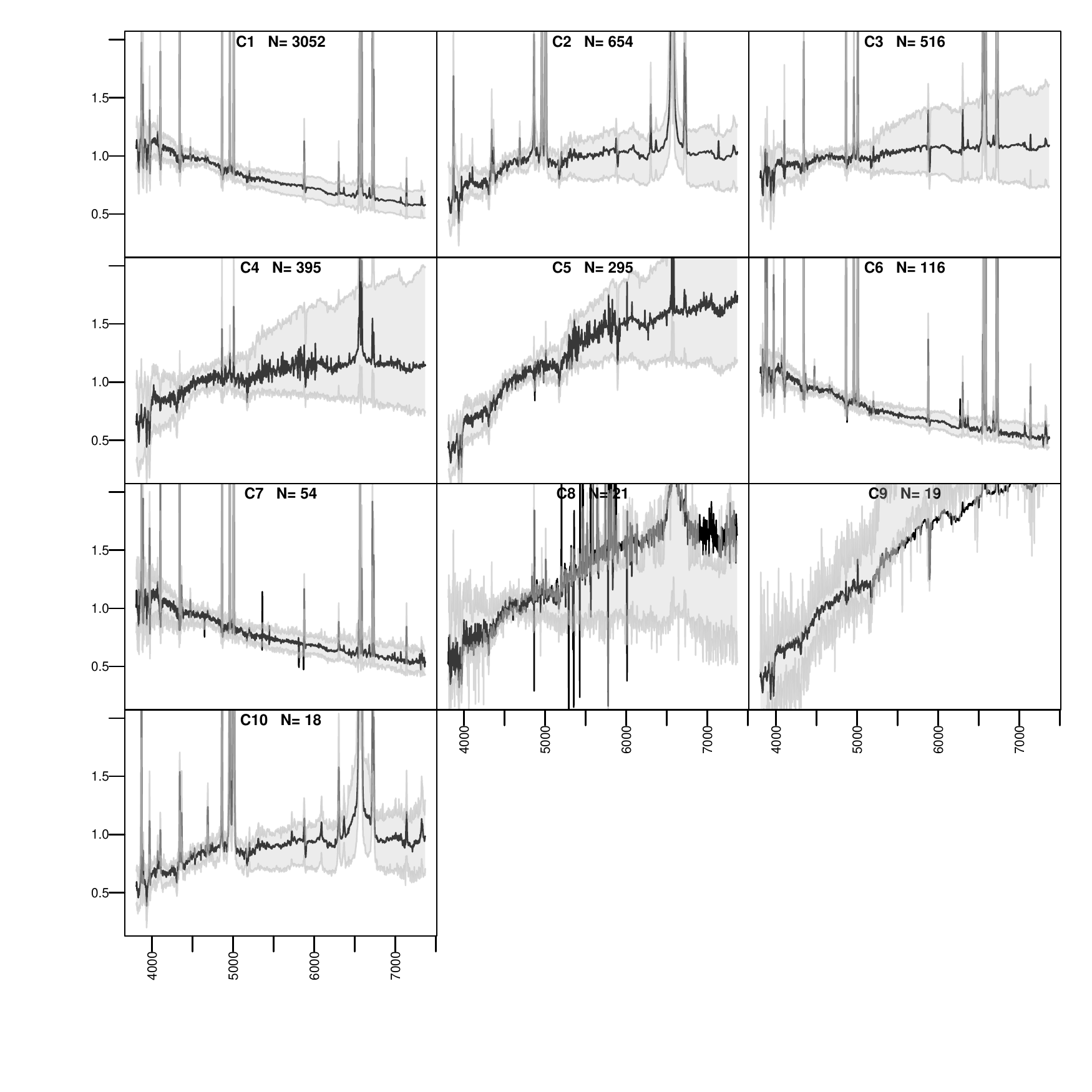}
\caption{Sub-classification of class C with 5140 spectra.}
\label{fig:ClassC}
\end{center}
\end{figure*}

The main classes A, B, and C were further analysed with Fisher-EM to find more refined clustering (Sects.~\ref{fourmainclasses}, \ref{subclassA}, \ref{subclassB} and \ref{subclassC}). We searched for the optimum number of classes using the ICL criterion for each of them independently (Figs.~\ref{fig:Aicl}, \ref{fig:Bicl}, \ref{fig:Cicl}). For class B there is a clear peak at K=25. For classes A and B, the ICL curve presents a strong increase at low Ks and then flattens significantly while continuing to increase slightly (by a few percent). We were not able to reach a peak and the results for larger K yielded more and more very small classes (1 to 3 members) while the larger classes do not change much. For instance in the case of class C, for K=60 the distribution is 2838, 703, 664, 416, 150, 90, 44, 41, 30, 20  ... 3 (8 classes), 2 (10) and 1 (20), so only 11 classes have more than 10 spectra, while for K=10 this distribution is 3052, 654, 516, 395, 295, 116, 54, 21, 19, 18. 
Because of this and because the value of ICL does not varie much, we adopted the value at the inflexion of the curves as shown by the red dot (i.e. 50 for A and 10 for C). 
The resulting classes are shown in Fig.~\ref{fig:ClassA}, \ref{fig:ClassB}, \ref{fig:ClassC}.

\clearpage

\section{Match with other atlases}
\label{app:otheratlases}

In this appendix, we present the comparison of our classification with some published atlases of templates as discussed in Sect.~\ref{comparison}.

\begin{figure*}[h]
\includegraphics[width=\linewidth]{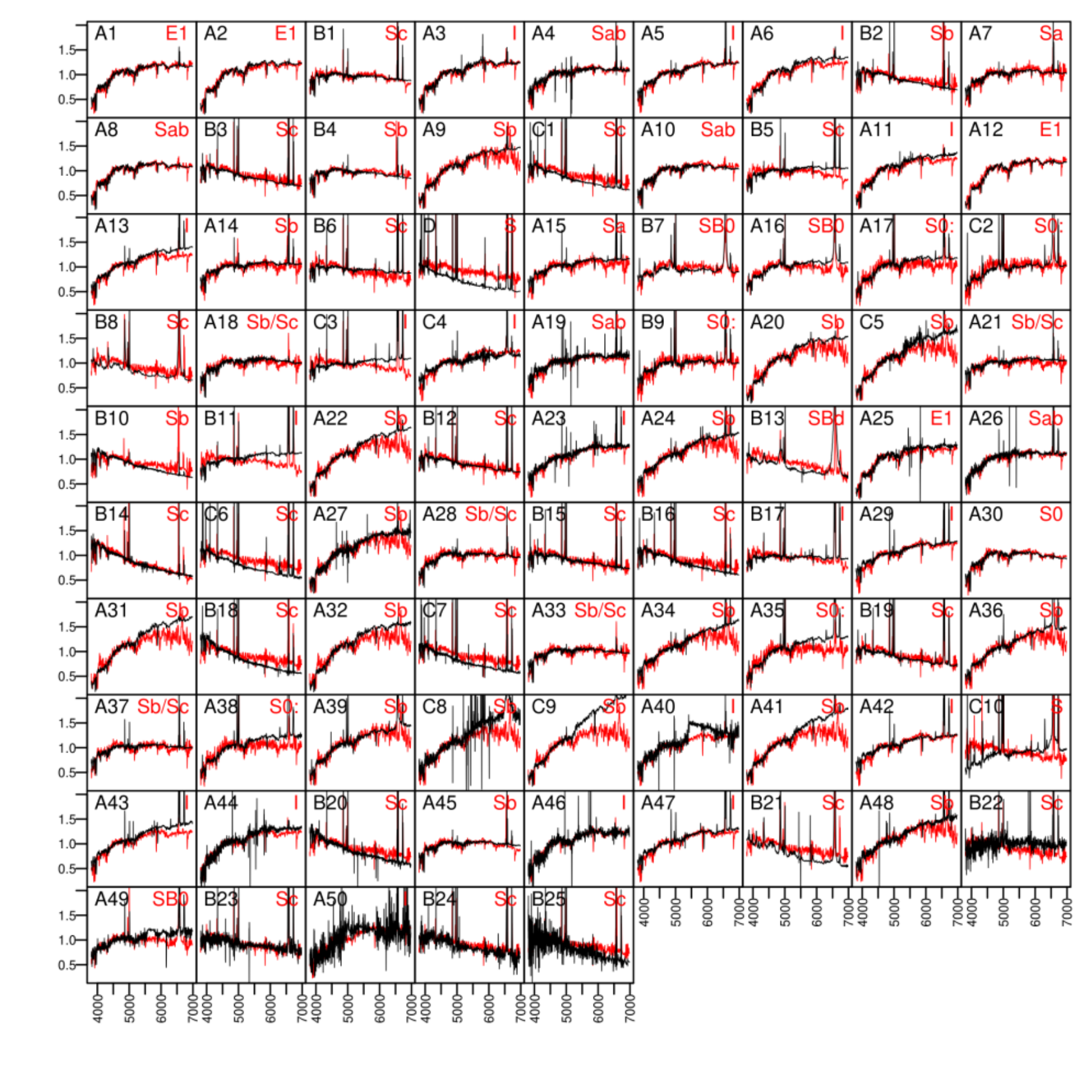}
\caption{kNN match of our classes to the atlas by \citet{Kennicutt1992}. See Sect.~\ref{comparison}.}
\label{fig:FEMKen}
\end{figure*}

\begin{figure*}[h]
\includegraphics[width=\linewidth]{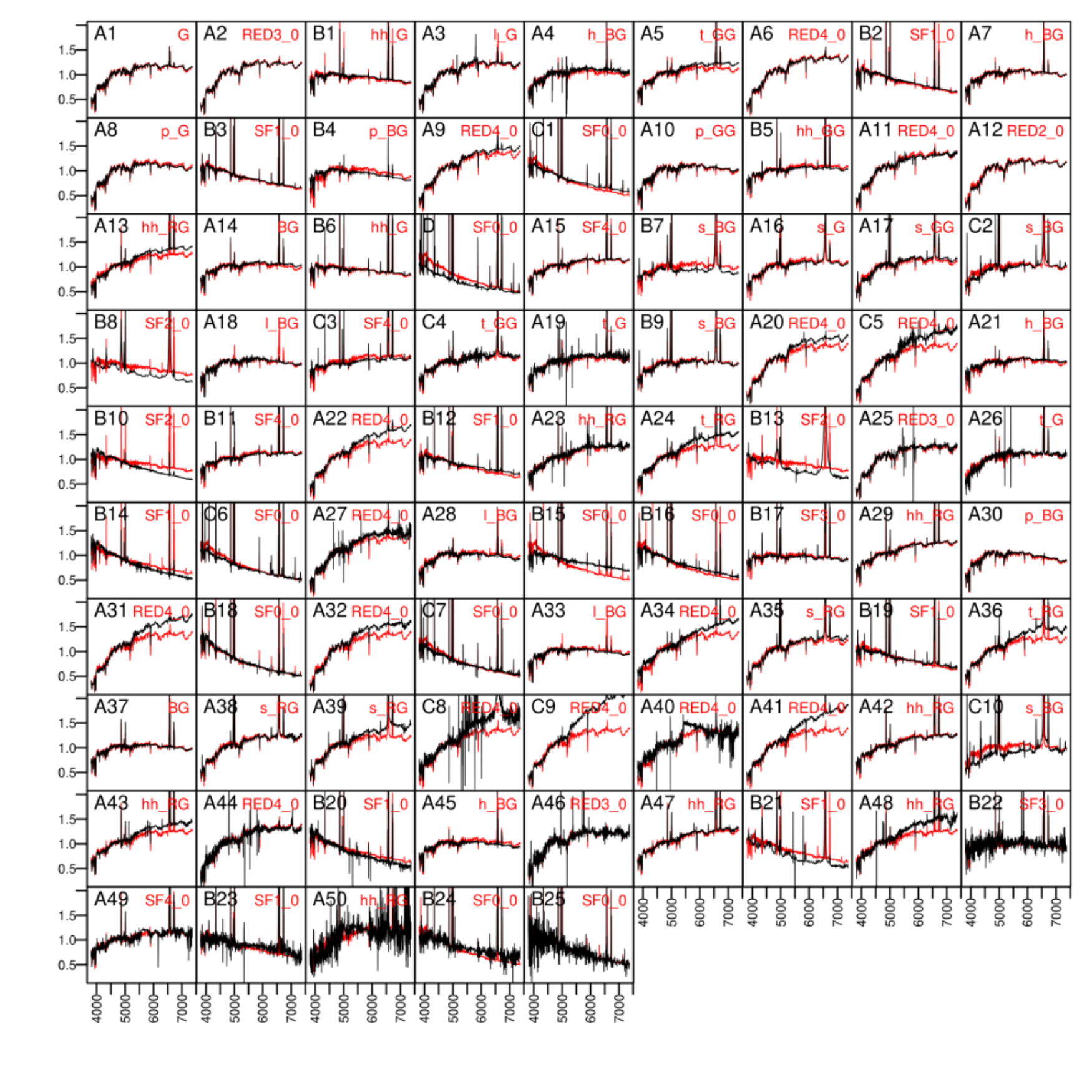}
\caption{Same as Fig.~\ref{fig:FEMKen} for the atlas by \citet{Dobos2012}.}
\label{fig:FEMDobos}
\end{figure*}

\begin{figure*}[h]
\includegraphics[width=\linewidth]{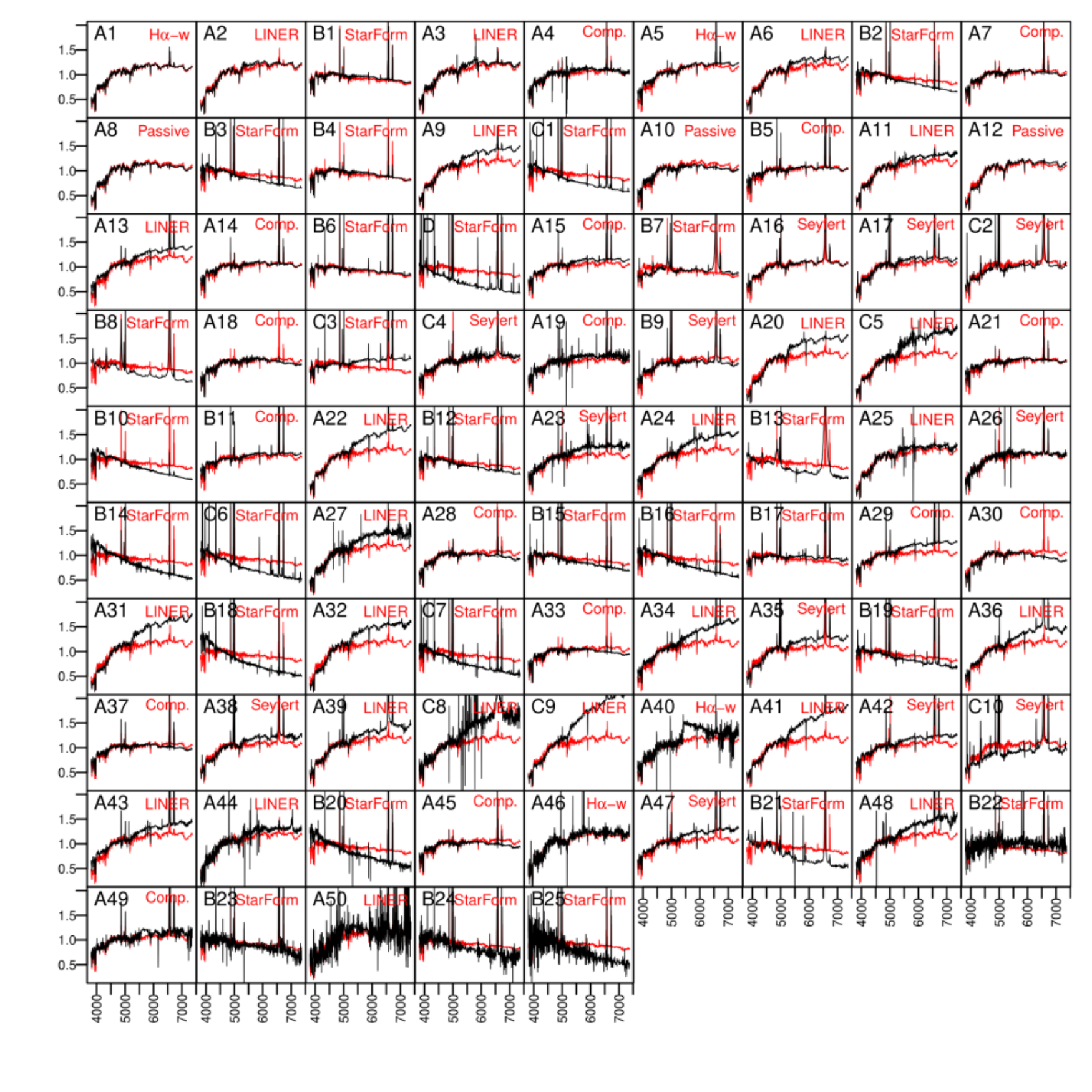}
\caption{Same as Fig.~\ref{fig:FEMKen} for the atlas by \citet{Wang2018}. "StarForm"} stands for Star Forming, H$\alpha$-w for H$\alpha$-weak and "Comp." for Composite galaxies.
\label{fig:FEMWang}
\end{figure*}

\begin{table*}[ht]
\centering
\begin{tabular}{llll|lllll}
\hline\hline
Fisher-EM & Kennicutt1992 &Dobos+2012 &Wang+2018 & &	Fisher-EM & Kennicutt1992 &Dobos+2012 &Wang+2018 \\
\hline
A1 & E1 & G & Halpha-weak &   & A25 & E1 & RED3\_0 & LINER \\ 
A2 & E1 & RED3\_0 & LINER &   & A26 & Sab & t\_G & Seyfert \\ 
B1 & Sc & hh\_G & Star Forming &   & B14 & Sc & SF1\_0 & Star Forming \\ 
A3 & I & l\_G & LINER &   & C6 & Sc & SF0\_0 & Star Forming \\ 
A4 & Sab & h\_BG & Composite &   & A27 & Sb & RED4\_0 & LINER \\ 
A5 & I & t\_GG & Halpha-weak &   & A28 & Sb/Sc & l\_BG & Composite \\ 
A6 & I & RED4\_0 & LINER &   & B15 & Sc & SF0\_0 & Star Forming \\ 
B2 & Sb & SF1\_0 & Star Forming &   & B16 & Sc & SF0\_0 & Star Forming \\ 
A7 & Sa & h\_BG & Composite &   & B17 & I & SF3\_0 & Star Forming \\ 
A8 & Sab & p\_G & Passive &   & A29 & I & hh\_RG & Composite \\ 
B3 & Sc & SF1\_0 & Star Forming &   & A30 & S0 & p\_BG & Composite \\ 
B4 & Sb & p\_BG & Star Forming &   & A31 & Sb & RED4\_0 & LINER \\ 
A9 & Sb & RED4\_0 & LINER &   & B18 & Sc & SF0\_0 & Star Forming \\ 
C1 & Sc & SF0\_0 & Star Forming &   & A32 & Sb & RED4\_0 & LINER \\ 
A10 & Sab & p\_GG & Passive &   & C7 & Sc & SF0\_0 & Star Forming \\ 
B5 & Sc & hh\_GG & Composite &   & A33 & Sb/Sc & l\_BG & Composite \\ 
A11 & I & RED4\_0 & LINER &   & A34 & Sb & RED4\_0 & LINER \\ 
A12 & E1 & RED2\_0 & Passive &   & A35 & S0: & s\_RG & Seyfert \\ 
A13 & I & hh\_RG & LINER &   & B19 & Sc & SF1\_0 & Star Forming \\ 
A14 & Sb & BG & Composite &   & A36 & Sb & t\_RG & LINER \\ 
B6 & Sc & hh\_G & Star Forming &   & A37 & Sb/Sc & BG & Composite \\ 
D & S & SF0\_0 & Star Forming &   & A38 & S0: & s\_RG & Seyfert \\ 
A15 & Sa & SF4\_0 & Composite &   & A39 & Sb & s\_RG & LINER \\ 
B7 & SB0 & s\_BG & Star Forming &   & C8 & Sb & RED4\_0 & LINER \\ 
A16 & SB0 & s\_G & Seyfert &   & C9 & Sb & RED4\_0 & LINER \\ 
A17 & S0: & s\_GG & Seyfert &   & A40 & I & RED4\_0 & Halpha-weak \\ 
C2 & S0: & s\_BG & Seyfert &   & A41 & Sb & RED4\_0 & LINER \\ 
B8 & Sc & SF2\_0 & Star Forming &   & A42 & I & hh\_RG & Seyfert \\ 
A18 & Sb/Sc & l\_BG & Composite &   & C10 & S & s\_BG & Seyfert \\ 
C3 & I & SF4\_0 & Star Forming &   & A43 & I & hh\_RG & LINER \\ 
C4 & I & t\_GG & Seyfert &   & A44 & I & RED4\_0 & LINER \\ 
A19 & Sab & t\_G & Composite &   & B20 & Sc & SF1\_0 & Star Forming \\ 
B9 & S0: & s\_BG & Seyfert &   & A45 & Sb & h\_BG & Composite \\ 
A20 & Sb & RED4\_0 & LINER &   & A46 & I & RED3\_0 & Halpha-weak \\ 
C5 & Sb & RED4\_0 & LINER &   & A47 & I & hh\_RG & Seyfert \\ 
A21 & Sb/Sc & h\_BG & Composite &   & B21 & Sc & SF1\_0 & Star Forming \\ 
B10 & Sb & SF2\_0 & Star Forming &   & A48 & Sb & hh\_RG & LINER \\ 
B11 & I & SF4\_0 & Composite &   & B22 & Sc & SF3\_0 & Star Forming \\ 
A22 & Sb & RED4\_0 & LINER &   & A49 & SB0 & SF4\_0 & Composite \\ 
B12 & Sc & SF1\_0 & Star Forming &   & B23 & Sc & SF1\_0 & Star Forming \\ 
A23 & I & hh\_RG & Seyfert &   & A50 & I & hh\_RG & LINER \\ 
A24 & Sb & t\_RG & LINER &   & B24 & Sc & SF0\_0 & Star Forming \\ 
B13 & SBd & SF2\_0 & Star Forming &   & B25 & Sc & SF0\_0 & Star Forming \\ 
\hline
\end{tabular}
\caption{Correspondence between our classes (Fisher-EM) and those of other atlases (\citet{Kennicutt1992, Dobos2012, Wang2018}). Our classes are ordered according to their number of spectra. \citet{Dobos2012} define the nomenclature as: p=passive, l=LINER, s=Seyfert, h=H$\alpha$, hh=AGN+HII, t=all, BG=all blue galaxies (idem for GG (green) and RG (red)).}
\label{tab:corresp}]
\end{table*}

\clearpage

\begin{figure*}[h]
\includegraphics[width=\linewidth]{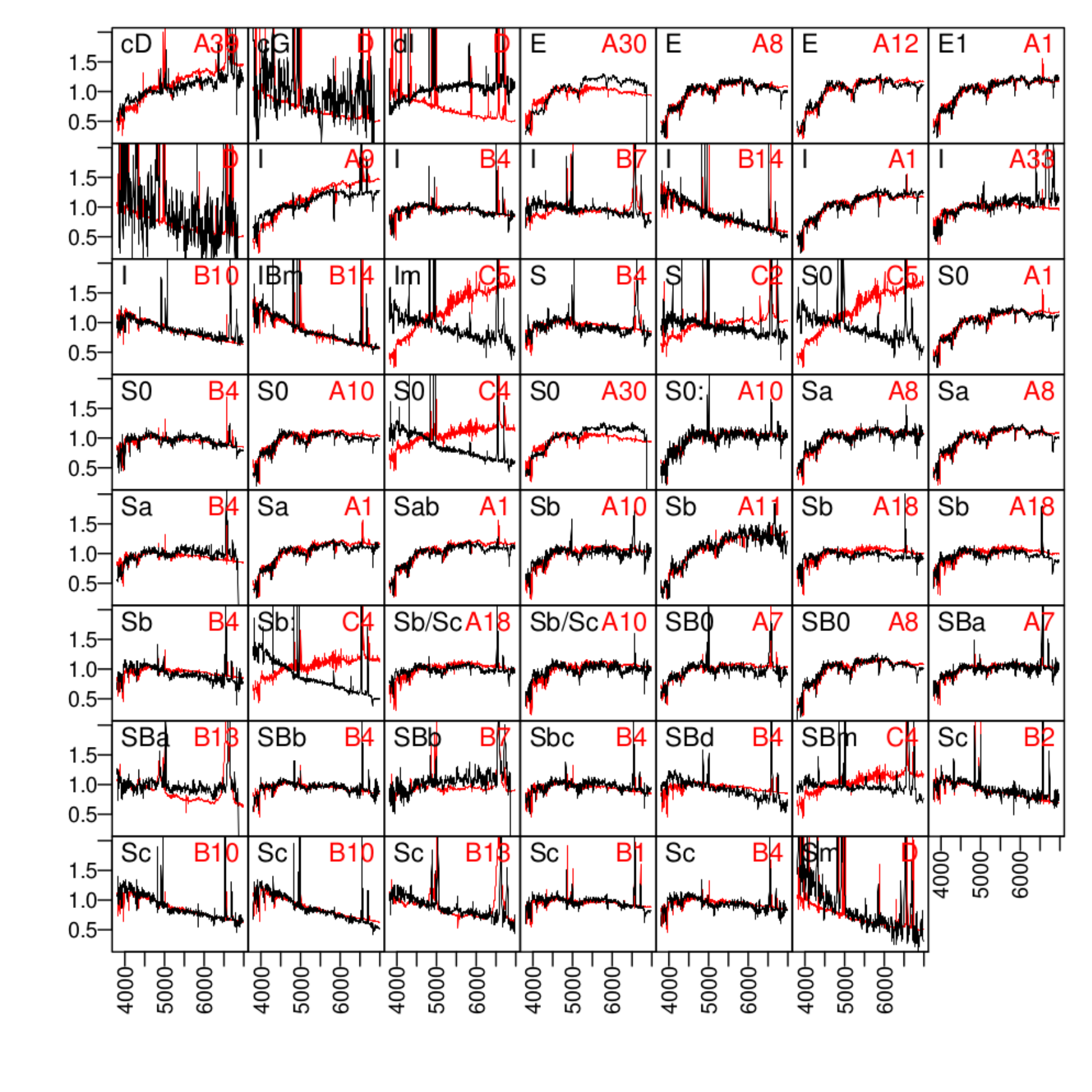}
\caption{Match of the atlas by \citet{Kennicutt1992} to our classes using the E-step. See Sect.~\ref{comparison}. The strong discrepancies in the fits for some classes (dI, Im, SO, Sb:) are due to very strong and thin emission lines in the \citet{Kennicutt1992} spectra. The fits are strongly improved when these spectra are smoothed to reduce the noise.}
\label{fig:KenFEM}
\end{figure*}

\begin{figure*}[h]
\includegraphics[width=\linewidth]{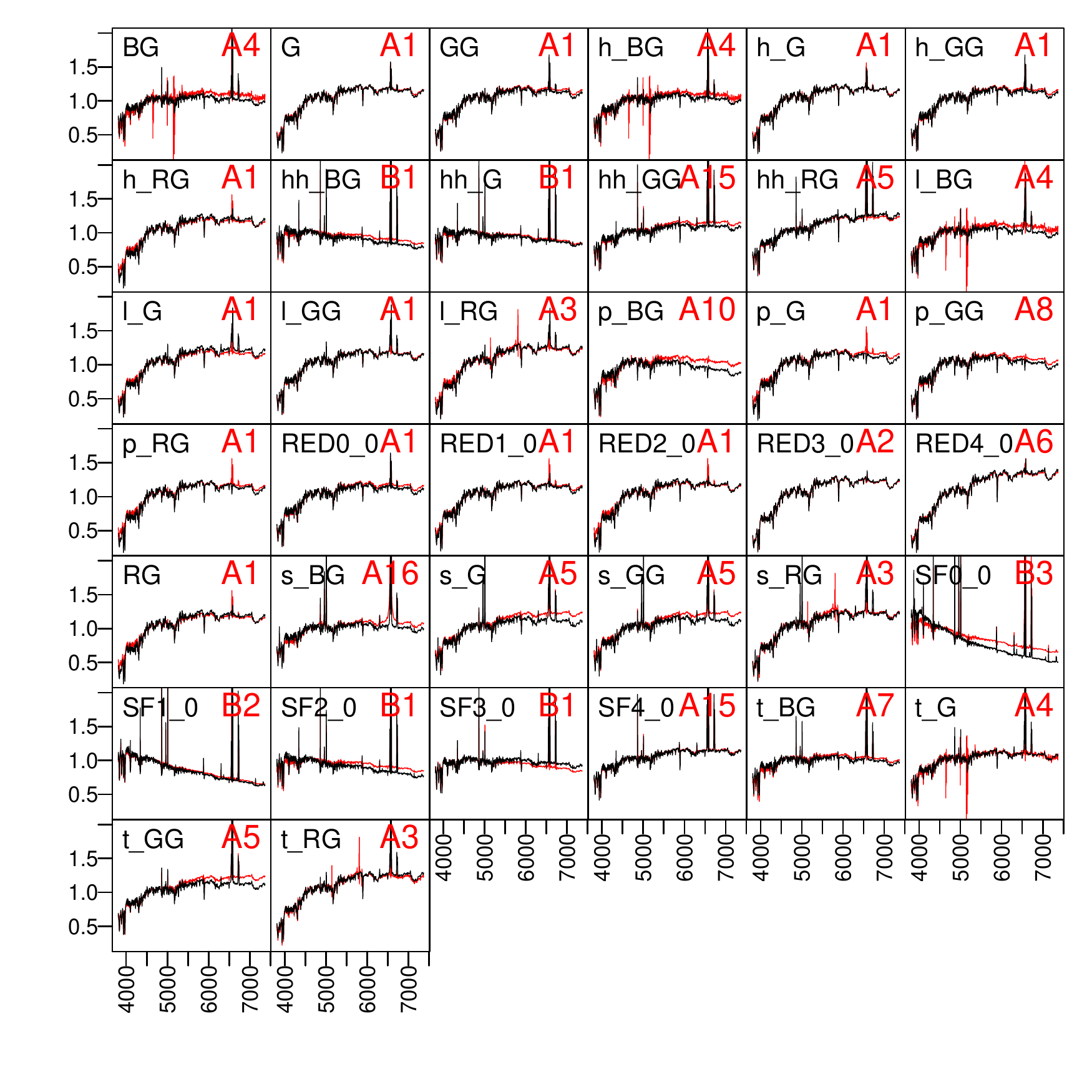}
\caption{Same as Fig.~\ref{fig:KenFEM} with the atlas by \citet{Dobos2012}.}
\label{fig:DobosFEM}
\end{figure*}

\begin{figure*}[h]
\includegraphics[width=\linewidth]{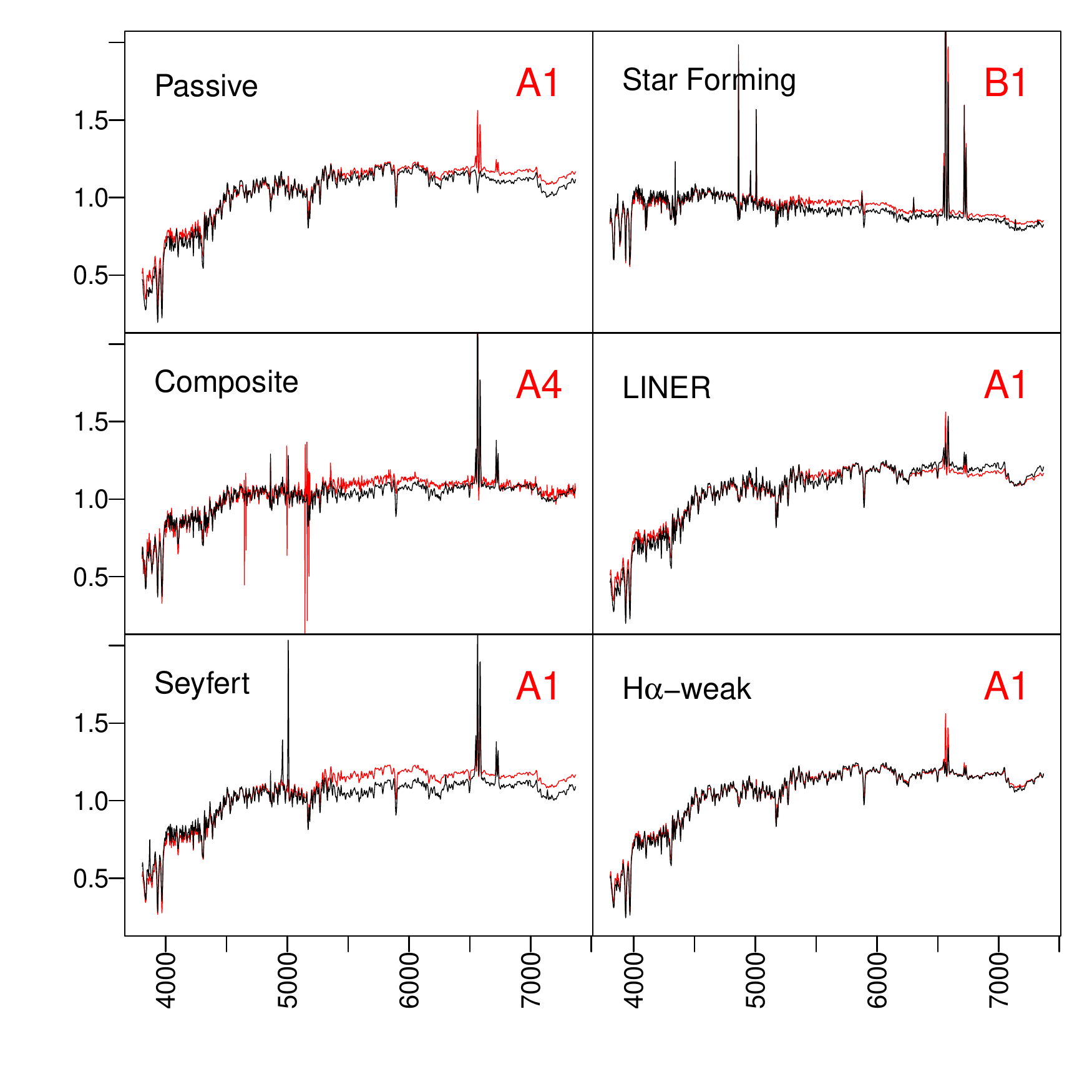}
\caption{Same as Fig.~\ref{fig:KenFEM} with the atlas by \citet{Wang2018}.}
\label{fig:WangFEM}
\end{figure*}

\end{appendix}

\end{document}